\documentclass[fleqn,usenatbib]{mnras}
\usepackage[T1]{fontenc}
\usepackage{graphicx}
\usepackage{amsmath}
\usepackage{amssymb}
\usepackage{xspace}
\usepackage{newtxtext,newtxmath}
\usepackage{multirow}
\usepackage[section]{placeins}

\newcommand{\nv}{\hat{\boldsymbol{\theta}}}
\newcommand{\rsfr}{\rho_{\rm SFR}}
\newcommand{\bsfr}{\langle b\rho_{\rm SFR}\rangle}

\newcommand{\nmt}{\texttt{NaMaster}\xspace}

\newcommand{\ccl}{\texttt{CCL}\xspace}
\newcommand{\cobaya}{\texttt{Cobaya}\xspace}
\newcommand{\planck}{{\sl Planck}\xspace}
\newcommand{\des}{DES\xspace}
\newcommand{\kids}{KiDS\xspace}

\defcitealias{2206.15394}{J22}
\newcommand{\JRGKA}{\citetalias{2206.15394}}
\defcitealias{2018MNRAS.477.1822M}{M18}
\newcommand{\MRG}{\citetalias{2018MNRAS.477.1822M}}

\DeclareRobustCommand{\VAN}[3]{#2}
\let\VANthebibliography\thebibliography
\def\thebibliography{\DeclareRobustCommand{\VAN}[3]{##3}\VANthebibliography}

\title[SFR constraints from CIB and cosmic shear]{Constraining the physics of star formation from CIB-cosmic shear cross-correlations}

\author[Jego et al.]{Baptiste Jego$^{1,2}$\thanks{E-mail: baptiste.jego@ens-paris-saclay.fr},
David Alonso$^{2}$,
Carlos Garc\'ia-Garc\'ia$^{2}$,
and Jaime Ruiz-Zapatero$^{2}$
\\
$^{1}$ENS Paris-Saclay, Gif-sur-Yvette, France\\
$^{2}$Department of Physics, University of Oxford, Denys Wilkinson Building, Keble Road, Oxford OX1 3RH, UK
}

\date{Accepted XXX. Received YYY; in original form ZZZ}

\pubyear{2022}

\begin{document}
\label{firstpage}
\pagerange{\pageref{firstpage}--\pageref{lastpage}}
\maketitle

% Abstract of the paper
\begin{abstract}
  Understanding the physics of star formation is one of the key problems facing modern astrophysics. The Cosmic Infrared Background (CIB), sourced by the emission from all dusty star-forming galaxies since the epoch of reionisation, is a complementary probe to study the star formation history, as well as an important extragalactic foreground for studies of the Cosmic Microwave Background (CMB). In this paper, we make high signal-to-noise measurements of the cross-correlation between maps of the CIB from the Planck experiment, and cosmic shear measurements from the Dark Energy Survey and Kilo-Degree Survey. Cosmic shear, is a direct tracer of the matter distribution, and thus we can use its cross-correlation with the CIB to directly test our understanding of the link between the star formation rate (SFR) density and the matter density. We use our measurements to place constraints on a halo-based model of the SFR that parametrises the efficiency with which gas is transformed into stars as a function of halo mass and redshift. These constraints are enhanced by using model-independent measurements of the bias-weighted SFR density extracted from the tomographic cross-correlation of galaxies and the CIB. We are able to place constraints on the peak efficiency at low redshifts, $\eta=0.445^{+0.055}_{-0.11}$, and on the halo mass at which this peak efficiency is achieved today $\log_{10}(M_1/M_\odot) = 12.17\pm0.25$. Our constraints are in excellent agreement with direct measurements of the SFR density, as well as other CIB-based studies.
\end{abstract}


\section{Introduction}\label{sec:intro}

  The study of the Cosmic Microwave Background (CMB) has evolved rapidly during the last two decades \citep{astro-ph/0302209,astro-ph/0603449,0803.0547,1001.4538,1009.0866,1212.5226,1301.0824,1303.5076,1502.01589,1702.03272,1807.06209,1910.07157,2007.07288}. Thanks to the swift advance in detector technology, and the construction of new ground-based facilities, we now have at our disposal high-sensitivity, wide-area maps of the radio and infrared sky at arc-minute resolutions. This situation will keep improving rapidly in the next few years with the advent of new facilities such as the Simons Observatory \citep{1808.07445}, the LiteBIRD satellite \citep{2101.12449}, and CMB S4 \citep{1610.02743}. However, through this gain in sensitivity, the field has transitioned to a regime where the contribution from various non-cosmological contaminants can no longer be ignored or avoided via masking, and must instead be modelled and incorporated into the cosmological analysis. This is particularly true for total intensity observations, where extra-galactic emission from radio sources, the thermal and kinematic Sunyaev-Zel'dovich (SZ) effects, and the Cosmic Infrared Background (CIB) \citep{1998ApJ...508..106D,1996A&A...308L...5P}, dominate the emission on small scales at all frequencies \citep{2011A&A...536A..18P,2013JCAP...07..025D}. Although the situation is less dire for polarisation data, whose constraining power has greatly increased, total intensity maps are still vital, as they can be used to constrain, for example, the epoch of reionisation, primordial non-Gaussianity, and the late-time growth of structure through the kinetic SZ effect \citep{1607.01769,1810.13424,1810.13423}.

  As a foreground contaminant, the CIB is particularly complex to treat. Its non-universal spectral energy distribution leads to significant decorrelation between observations at different frequencies, which complicates its removal through the simplest multi-frequency component separation methods. As an indirect tracer of the large-scale matter fluctuations peaking at redshift $z\sim 2$, it has a markedly non-Gaussian structure that can contaminate the reconstruction of the CMB lensing potential, with which it correlates strongly \citep{2014JCAP...03..024O,2014ApJ...786...13V,2021PhRvD.104l3514S,2021arXiv211100462D}. Understanding the physics of the CIB is therefore of vital importance for CMB cosmology, in order to devise more effective component separation techniques, and to incorporate the residual contamination in the model used to obtain cosmological constraints.
  
  In spite of its role as a nuisance in CMB observations, the CIB is a remarkable tool for astrophysics. In cosmology, it can be used as a probe of structure on ultra-large scales \citep{1606.02323}, and to revert the effects of gravitational lensing on maps of the CMB \citep{1010.0048,1502.05356}. More importantly, the CIB contains invaluable information to improve our understanding of the formation and evolution of galaxies. The widely accepted origin of the CIB is the combined infrared emission from dust in star-forming galaxies, heated by the absorption of ultraviolet (UV) light from massive short-lived stars \citep{1967ApJ...147..868P,2001ApJ...550....7K}. As such, maps of the CIB contain information about the star formation rate (SFR) history from the epoch reionisation until today \citep{2006A&A...451..417D}. Understanding the formation of stars in galaxies of different types at different cosmic epochs is key to understand the formation and evolution of galaxies themselves \citep{1980FCPh....5..287T}. Studies of the SFR history have evolved rapidly thanks to the observation of the UV and infrared luminosity function \citep{2013MNRAS.432...23G,2013A&A...553A.132M,2016MNRAS.456.1999M,2016MNRAS.461..458D}. These studies have shown that the star formation rate density (SFRD) grows swiftly from the epoch of reionisation, peaking at $z\sim2$, and then decreasing as the gas fueling it is depleted \citep{2014ARA&A..52..415M}. 

  Although the global picture is qualitatively well understood, the details of the relation between SFR and galaxy properties, or the properties of the halos these galaxies reside in, is far murkier. Studies of the CIB are able to shed some light. First studying the CIB anisotropies through their auto-correlation can constrain the spatial distribution of infrared sources, although the projected nature of the CIB maps makes it difficult to disentangle the contributions from different redshifts \citep{2012MNRAS.421.2832S,2013ApJ...772...77V,2014A&A...571A..30P,1801.10146,2021A&A...645A..40M}. This can be remedied through cross-correlations. \citet{2206.15394} (J22 hereafter) recently showed that the cross-correlation between the CIB and a set of galaxy samples on large scales can be used to make a model-independent tomographic measurement of the bias-weighted SFR density $\bsfr$ which, when combined with direct measurements of the SFRD $\rsfr$, can shed light on the relation between SFR and halo mass as a function of redshift. The same cross-correlation on small scales is, in principle, sensitive to the relation between SFR and the properties of the target galaxies \citep{2014A&A...570A..98S,2015MNRAS.449.4476W,2016ApJ...831...91C,2022arXiv220401649Y}. In this regime the signal is sensitive to the fraction of star-forming galaxies in the target sample, which complicates its modelling and interpretation.

  In this paper, we will turn instead to correlations between the CIB and tomographic galaxy weak lensing measurements. Since the cosmic shear signal from galaxies at different redshifts directly traces matter inhomogeneities, this cross-correlation is sensitive to the relation between the SFR and matter densities at different times. This will thus allow us to test the validity of different halo-based SFR models and, in general, to improve our understanding of the connection between star-forming galaxies and the underlying dark matter fluctuations. This approach is complementary to the study of the correlation between the CIB and the lensing convergence of the CMB \citep{2014A&A...571A..18P,1801.10146,2020ApJ...901...34C,2021MNRAS.500.2250D,2021PhRvD.103j3515M} with two advantages. First, cosmic shear data provide a handle on the redshift dependence of the signal via tomography. Secondly, while the CIB is a known contaminant for CMB lensing reconstruction, no such contamination exists for cosmic shear.

  This paper is structured as follows. Section \ref{sec:methods} presents the theoretical background and the methods used in the analysis. The datasets used are described in Section \ref{sec:data}. The measured cross-correlations are presented and analysed in Section \ref{sec:results}, where we also present the associate constraints on star formation models. We summarise and discuss our results in Section \ref{sec:conc}.

\section{Methods}\label{sec:methods}
  \subsection{Theory}\label{ssec:methods.theo}
    Our theory prediction will follow the formalism described in \citep{2018MNRAS.477.1822M,2021JCAP...10..030G,2206.15394}.

    \subsubsection{Angular power spectra and the halo model}\label{sssec:methods.theo.clhm}
      The cosmic shear signal and CIB anisotropies can both be described in as projected tracers (i.e. sky maps) $u(\nv)$ of a three-dimensional field $U({\bf x},z)$ through a radial kernel $q_u(\chi)$:
      \begin{equation}\label{eq:projected.anisotropy}
        u(\nv) = \int d\chi q_u(\chi)\,U(\chi\nv,z),
      \end{equation}
      where $z$ is the redshift corresponding to the comoving distance $\chi$. The angular power spectrum of two such quantities, $u$ and $v$ is then given by
      \begin{equation}\label{eq:angular.c_ell}
        C_\ell^{uv} = f_\ell^uf_\ell^v\int \frac{d\chi}{\chi^2} q_u(\chi)q_v(\chi) P_{UV}\left(k = \frac{\ell + 1/2}{\chi},z\right),
      \end{equation}
      where $P_{UV}(k,z)$ is the power spectrum of the corresponding 3D quantities, and we have made use of the Limber approximation \citep{1953ApJ...117..134L}, which is appropriate for the tracers analysed here. The multiplicative factors $f^{u/v}_\ell$ account for the potential angular derivatives relating the 2D and 3D quantities.

      We will model $P_{UV}(k,z)$ making use of the halo model \citep{astro-ph/0001493,astro-ph/0005010,astro-ph/0206508}. In this formalism
      \begin{equation}
        P_{UV}(k,z)=P^{2h}_{UV}(k,z)+P^{1h}_{UV}(k,z),
      \end{equation}
      where the 1-halo and 2-halo contributions are given by
      \begin{align}\label{eq:1-halo}
        &P^{1h}_{UV}(k) \equiv \int_{}^{}dM\, n(M) \langle U(k,M)V(k,M)\rangle,\\\label{eq:2-halo}
        &P^{2h}_{UV}(k) \equiv \langle bU(k)\rangle \langle bV(k)\rangle P_{\rm lin}(k),\\\label{eq:bU}
        &\langle bU(k)\rangle\equiv\int dM\,n(M)\,b_h(M)\,\langle U(k,M)\rangle.
      \end{align}
      Here, $n(M)$ and $b_h(M)$ are the halo mass function and the halo bias respectively for halos of mass $M$, and $P_{\rm lin}(k,z)$ is the linear matter power spectrum. $\langle U(k,M)\rangle$ is the Fourier transform of the mean halo profile of quantity $U$, i.e.
      \begin{equation}
        \langle U(k,M)\rangle=4\pi\int_0^\infty dr\,r^2\,\langle U(r,M)\rangle\,\frac{\sin kr}{kr},
      \end{equation}
      where $\langle U(r,M)\rangle$ is the mean value of $U$ at a distance $r$ from the center of a halo of mass $M$. Likewise, $\langle U(k,M)V(k,M)\rangle$ is the two-point cumulant of the two Fourier-space profiles.

      A prediction for the angular cross-power spectrum between cosmic shear (labelled as $\gamma$ here) and CIB anisotropies at frequency $\nu$ thus requires the radial kernels of both probes, and a model for the statistics (scale-dependent mean and covariance) of the 3D quantities associated with them in halos of different masses. We describe these ingredients in the next two sections.

      Although the halo model is able to describe the power spectrum on either large- or small-scales, where either the 2-halo or the 1-halo terms dominate, it is inaccurate at the $\sim10-20\%$ level on intermediate scales. The most likely cause for this is the oversimplified treatment of halo biasing used in the vanilla version of the halo model used here \citep{2021MNRAS.502.1401M,2021MNRAS.503.3095M}. We account for this following \citep{1606.05345,1909.09102}. We multiply the halo model prediction for $P_{\gamma\nu}(k,z)$ in the range $0.05\,{\rm Mpc}^{-1}<k<2\,{\rm Mpc}^{-1}$ by the factor
      \begin{equation}
        R(k,z)\equiv\frac{P_{\rm halofit}(k,z)}{P_{\rm HM}(k,z)},
      \end{equation}
      where $P_{\rm HM}(k,z)$ is the halo model prediction for the matter-matter power spectrum, and $P_{\rm halofit}(k,z)$ is the fit to the same quantity using the {\tt Halofit} \citep{astro-ph/0207664} parametrisation of \citet{1208.2701}.

    \subsubsection{Cosmic shear and matter fluctuations}\label{sssec:methods.theo.wl}
      Weak gravitational lensing distorts the shapes of background galaxies, correlating their ellipticities. This effect, known as ``cosmic shear'' is quantified through a spin-2 projected field $\gamma$. At leading order, weak lensing contributes only to the parity-even ``$E$-mode'' component of the field. This is a projected tracer of the matter overdensity $\Delta_m({\bf x},z)$ with radial kernel \citep{astro-ph/9912508}
      \begin{equation}\label{eq:kernel_wl}
        q_\gamma(\chi)=\frac{3}{2}H_0^2\Omega_m(1+z)\chi\int_z^\infty dz'\,p(z')\,\frac{\chi(z')-\chi}{\chi(z')},
      \end{equation}
      where $H_0$ is the current value of the expansion rate, $\Omega_m$ is the fractional energy density of non-relativistic matter, and $p(z)$ is the redshift distribution of the source galaxies. We use natural units where the speed of light is $c=1$.

      The $\ell$-dependent prefactor, due to the relation between $\gamma$ and the angular Hessian of the Newtonian gravitational potential, is
      \begin{equation}
        f_\ell^\gamma\equiv\sqrt{\frac{(\ell+2)!}{(\ell-2)!}}\frac{1}{(\ell+1/2)^2},
      \end{equation}
      which is negligibly different from 1 on the scales used here.

      The associated halo profile is simply the matter density profile normalised by the mean background matter density $\bar{\rho}_M$. For this we use the truncated Navarro-Frenk-White parametrisation \citep{astro-ph/9508025}. In this case
      \begin{equation}
        \langle u_\gamma(k,M)\rangle = \frac{M}{\bar{\rho}_m}u_{\rm NFW}(k,M),
      \end{equation}
      where
      \begin{align}\nonumber
        u_{\rm NFW}(k)=\left[\ln(1+c)-\frac{c}{1+c}\right]^{-1}\biggl\{\sin x\left[{\rm Si}((1+c)x)-{\rm Si}(x)\right]+ \\
        \left.\cos x\left[{\rm Ci}((1+c)x)-{\rm Ci}(x)\right]-\frac{\sin(cx)}{(1+c)x}\right\}.
      \end{align}
      Here $x\equiv k R_\Delta/c(M)$, $R_\Delta(M)$ is the halo virial radius, $c(M)$ is the concentration-mass relation, and ${\rm Si}/{\rm Ci}$ are the sine and cosine integrals.

      The intrinsic alignment (IA) of the shapes of galaxies, caused by local gravitational tidal forces, is an important source of contamination for cosmic shear \citep{astro-ph/0009499}. Unlike the cosmic shear signal, IAs are a local effect and correlate with the structure at redshift of the source galaxies. Within the linear non-linear alignment model (LNLA, \citet{astro-ph/0406275}), the effect of IAs can be taken into account by adding a local contribution to the cosmic shear kernel of the form:
      \begin{equation}
        q_I(\chi)=-A_{\rm IA}(z)\,H(z)\,p(z),
      \end{equation}
      where $H(z)$ is the expansion rate at redshift $z$, and $A_{\rm IA}$ is a linear amplitude parameter that parametrises the strength with which local tidal forces modify galaxy ellipticities. A common parametrisation of this amplitude is \citep[e.g.][]{1708.01538}
      \begin{equation}
        A_{\rm IA}(z)=A_0\left(\frac{1+z}{1+z_0}\right)^\lambda\,\frac{0.0139\Omega_m}{D(z)},
      \end{equation}
      where $D(z)$ is the linear growth factor, $z_0$ is a fixed pivot redshift, and $A_0$ and $\lambda$ are free parameters.

      We must also propagate uncertainties due to the poor knowledge of the source redshift distribution $p(z)$ of the shear samples used here. We do so by marginalising over a nuisance parameter $\Delta z$ in each redshift bin describing a shift in the mean of the redshift distribution:
      \begin{equation}
        p(z)\,\rightarrow\,p(z+\Delta z).
      \end{equation}
      This has been shown to encapsulate the main effect of redshift uncertainties on the weak lensing kernel \citep{2020OJAp....3E...6T,2022MNRAS.511.2170C}.

    \subsubsection{CIB and star formation}\label{sssec:methods.theo.cib_sfr}
      To model the CIB signal we follow the prescription of \cite{2018MNRAS.477.1822M,1801.10146,2021A&A...645A..40M,2206.15394} (the reader is referred to these papers for further details to avoid repetition).
      
      The CIB specific intensity at a given observed frequency $\nu$, $I_\nu(\nv)$, is a projected tracer of the SFRD, $\rsfr$:
      \begin{equation}\label{eq:cib_proj}
        I_\nu(\nv)=\int d\chi\,q_\nu(\chi)\,\rsfr(\chi\nv,z),
      \end{equation}
      with a radial kernel given by
      \begin{equation}\label{eq:kernel_cib}
        q_\nu(\chi)=\frac{\chi^2S_\nu^{\rm eff}(z)}{K}.
      \end{equation}
      Here, $K=10^{-10}\,M_\odot\,{\rm yr}^{-1}L_\odot^{-1}$ is the calibration constant relating the far infrared luminosity $L_{\rm IR}$ and star formation rate \citep[$L_{\rm IR}={\rm SFR}/K$][]{1998ARA&A..36..189K,2012ARA&A..50..531K} for a Chabrier initial mass function \citep{2003PASP..115..763C}, and $S^{\rm eff}_\nu(z) \equiv S_\nu(z) / L_{\rm IR}$ is the normalised mean spectral energy distribution of sources as a function of redshift\footnote{We use the estimates of \citet{2013A&A...557A..66B,2015A&A...573A.113B,2017A&A...607A..89B} for the \planck channels, made available in \url{https://github.com/abhimaniyar/halomodel_cib_tsz_cibxtsz}.}. The $\ell$-dependent factor is simply $f_\ell^\nu=1$.

      To model the SFRD profile, we add the contribution from central and satellite galaxies. The resulting mean Fourier-space profile is
      \begin{equation}
        \langle u_{\rm SFR}(k,M)\rangle={\rm SFR}_c(M,z)+{\rm SFR}_s(M,z)u_s(k,M),
      \end{equation}
      where ${\rm SFR}_{c/s}(M,z)$ is the star formation from centrals and satellites in a parent halo of mass $M$ at redshift $z$, and $u_s(k,M)$ is the mean distribution of satellites in Fourier space (normalised to $u_s(k\rightarrow0,M)=1$). For simplicity, we assume that satellites follow the dark matter distribution, and use $u_s(k,M)=u_{\rm NFW}(k,M)$. Note that this is an assumption that can be explicitly tested through the CIB-cosmic shear cross-correlation on small, halo-sized scales, although we do have not attempted to do so here.

      As in \cite{1801.10146,2021A&A...645A..40M,2018MNRAS.477.1822M}, we parametrise the SFR of the central galaxy in terms of the efficiency $\eta$ to convert infalling gas into stars:
      \begin{equation}
        {\rm SFR}_c(M,z)=\eta(M,z)\,{\rm BAR}(M,z),
      \end{equation}
      where ${\rm BAR}$ is the baryonic accretion rate, which we model after \citet{2010MNRAS.406.2267F} as
      \begin{equation}
        {\rm BAR}(M,z)=\dot{M}_0\frac{\Omega_b}{\Omega_M}\,\left(\frac{M}{10^{12}M_\odot}\right)^{1.1}(1+1.11z)\frac{H(z)}{H_0},
      \end{equation}
      with $\dot{M}_0=46.1 M_\odot\,{\rm yr}^{-1}$.
      
      To calculate the satellite contribution, we will assume that all subhalos in a parent halo of mass $M$ contain a satellite:
      \begin{equation}
        {\rm SFR}_s(M,z)=\int_{M_{\rm min}}^M dM_{\rm sub}\,\frac{dN}{dM_{\rm sub}}\,{\rm SFR}_{\rm sat}(M_{\rm sub},M,z),
      \end{equation}
      where $dN/dM_{\rm sub}$ is the subhalo mass function, parametrised as in \cite{2010ApJ...719...88T}. ${\rm SFR}_{\rm sat}(M_{\rm sub},M,z)$ is the SFR of a satellite galaxy with subhalo mass $M_{\rm sub}$. As in \cite{2021A&A...645A..40M}, we model ${\rm SFR}_{\rm sat}$ as
      \begin{equation}\nonumber
        {\rm SFR}_{\rm sat}(M_{\rm sub},M,z)={\rm Min}\left[{\rm SFR}_c(M_{\rm sub},z),\frac{M_{\rm sub}}{M}{\rm SFR}_c(M,z)\right].
      \end{equation}
      This guarantees that the SFR in any satellite never exceeds that of the central galaxy \citep[see][]{2021A&A...645A..40M}.
      
      Most parametrisations of $\eta(M,z)$ assume that the efficiency peaks at a particular halo mass $M_{\rm max}\sim10^{12-13}\,M_\odot$. This behaviour is physically motivated \citep{2003MNRAS.343..249S,2005MNRAS.363....2K}: at lower masses, the weaker gravitational potential and the impact of supernova feedback deplete the galaxy of gas. In turn, at higher masses the increased gas cooling time and the impact of AGN feeback have a similar effect. Here, we will choose the parametrisation of \citet{2018MNRAS.477.1822M} (M18 hereafter), which assumes
      \begin{equation}
        \eta(M,z)=\frac{2\eta_*}{(M_1/M)^\beta+(M/M_1)^\gamma}.
      \end{equation}
      All parameters ($\eta_*$, $m\equiv\log_{10}M_1/M_\odot$, $\beta$, and $\gamma$) are allowed to vary with redshift as:
      \begin{equation}
        x(z)=x_0+x_z\frac{z}{1+z}.
      \end{equation}
      As in \JRGKA{}, we will fix $\beta_0,\,\beta_z,\,\gamma_0$, and $\gamma_z$ to the best-fit values found by \MRG{}, and we will only consider four free parameters: $\eta_0,\,\eta_z,\,m_0$, and $m_z$. These parametrise the value and time dependence of the peak efficiency and the associated mass. Note that other models have been proposed in the literature. For instance, \citet{2021A&A...645A..40M}, proposed using a log-normal form for the mass dependence of $\eta(M,z)$, with a constant peak efficiency and mass, but a redshift-dependent high-mass tail. We choose to follow the parametrisation of \MRG{} here, since it arguably allows for a more flexible, but simple, redshift evolution of the efficiency curve.

      Given a halo-based model for the SFR, the mean star formation rate density can be calculated by simply integrating over the halo mass function:
      \begin{equation}
        \rsfr(z)=\int dM\,n(M,z)\,{\rm SFR}(M,z).
      \end{equation}
      Likewise, the bias-weighted version of the same quantity is
      \begin{equation}\label{eq:bsfr}
        \bsfr \equiv \int dM\,n(M,z)\,b_h(M,z)\,{\rm SFR}(M,z).
      \end{equation}

  \subsection{Angular power spectra and covariances}\label{ssec:methods.map2cl}
    \begin{figure}
      \centering
      \includegraphics[width=0.5\textwidth]{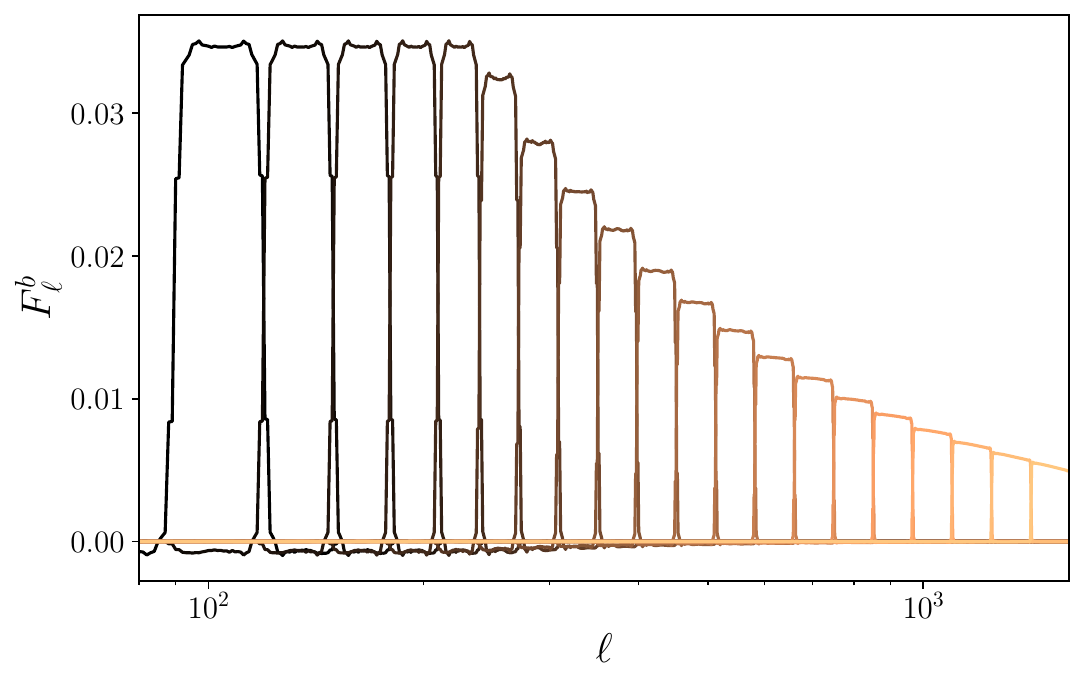}
      \caption{Bandpower window functions for the cross-correlation between the third \kids redshift bin and the 545 GHz CIB map.}
      \label{fig:bpw}
    \end{figure}
    We use the MASTER algorithm \citep{astro-ph/0105302} as implemented in \nmt \citep{1809.09603} to estimate all power spectra used in this analysis. The method is based on estimating the statistical coupling between different power spectrum multipoles, caused by the survey mask, (the so-called mode-coupling matrix) using analytical approach enabled by the orthogonality of the Wigner-3$j$ symbols. A thorough description of the method can be found in \citet{1809.09603}. Here we will only describe some of the details specific to the use of cosmic shear data.
    
    We follow the prescription of \citet{2010.09717}. The cosmic shear field $\gamma$ is only sampled at the positions of the observed galaxies. As such, the natural survey mask that optimises the pixel-level signal-to-noise is given by the sum of shape measurement weights for all galaxies in each pixel. This leads to a complex mask with substantial structure, which in turn can cause a non-negligible statistical coupling between distant multipoles. When comparing the estimated power spectra with theoretical predictions we therefore convolve the predictions with the bandpower window functions, which fully describe this coupling:
    \begin{equation}
      C_b = \sum_\ell {\cal F}_b^\ell\,C_\ell,
    \end{equation}
    where $C_\ell$ is the per-$\ell$ theoretical power spectrum, ${\cal F}_b^\ell$ is the window function for the $b$-th bandpower, and $C_b$ is the corresponding prediction for that bandpower. To illustrate this, Fig. \ref{fig:bpw} shows the bandpower windows for the cross-correlation between the third \kids bin and the 545 GHz CIB map.

    We will only include angular scales $100\leq\ell\leq1500$ in our analysis. The large-scale cut is motivated by the loss of power on large scales in the CIB maps due to Galactic dust removal, while the small-scale cut ensures that our analysis is not significantly affected by modelling uncertainties on the impact of baryonic feedback processes on the matter power spectrum. Using the same $\ell$ bins as \citep{2021JCAP...10..030G}, this leaves 20 bandpower measurements in for each cross correlation.

    We use the analytical approximation of \citet{1906.11765,2010.09717} (the so-called ``Narrow-Kernel Approximation'') to estimate the Gaussian covariance matrix of the data, neglecting all non-Gaussian contributions. The approximation relies on an estimate of the angular power spectra of the different tracers involved. For this, we use the pseudo-$C_\ell$ estimate from the data for all pairs of maps, normalised by the mean of the product of their respective masks. This has the advantage of not relying on a particular astrophysical model for the signal, or a precise instrument model for the noise. We validate this estimate by recomputing the covariance for a subset of the spectra via jackknife resampling. As in \JRGKA{}, we find that the analytical approximation works well at the per-cent level, although a small correction of a few percent must be applied to covariance matrix elements involving two different CIB maps. This is important in a multi-frequency analysis given the very high correlation between the different frequencies. Since, as we will describe in Section \ref{sssec:data.CIBLenz.coadd}, we will instead combine all frequencies at the map level for this analysis, this modification is less critical than in \JRGKA{}.

  \subsection{Likelihood}\label{ssec:methods.like}
      \begin{table}
        \centering
        \def\arraystretch{1.2}
        \begin{tabular}{llll}
        \hline
        \hline
        Parameter &  Prior & Parameter &  Prior\\  
        \hline
        $m_0$ & $U(9, 14)$ & $m_z$ & $U(-6,6)$ \\
        $\eta_0$ & $U(0,1)$ & $\eta_z$ & $U(0,1)$ \\
        $A_0^{\rm DES}$ & ${\cal N}(1,0.5)$ & $A_0^{\rm KiDS}$ & ${\cal N}(1,0.5)$ \\
        $\Delta z_{\rm DES}\,(\times4)$ & ${\cal N}(\boldsymbol{\mu},{\sf C})$ & $\Delta z_{\rm KiDS}\,(\times5)$ & ${\cal N}(\boldsymbol{\mu},{\sf C})$\\
        \hline
        \hline
        \end{tabular}
        \caption{Prior distributions for the nuisance parameters entering our analysis for each tracer. $U(a, b)$ and ${\cal N}(\boldsymbol{\mu}, {\sf C})$ describe a uniform distribution with boundaries $(a, b)$ and a multivariate Gaussian distribution with mean $\boldsymbol{\mu}$ and covariance ${\sf C}$, respectively.}\label{tab:priors}
      \end{table}
    We will make use of a Gaussian likelihood, 
    \begin{equation}\label{eq:like}
      -2\log p({\bf d}|\Theta)=({\bf d}-{\bf m}(\Theta))^T{\sf C}_{\bf d}^{-1}({\bf d}-{\bf m}(\Theta))+K,
    \end{equation} 
    to constrain the free parameters of the star-formation model. In this expression ${\bf d}$ is the data vector, containing all cross-correlations between tomographic cosmic shear bins and CIB maps, ${\sf C}_{\bf d}$ is the covariance of ${\bf d}$, ${\bf m}$ is the theoretical model for ${\bf d}$, and $\Theta$ denotes the free parameters of the model. The posterior distribution is then given by the product of this likelihood and the parameter priors.

    Our fiducial model contains 4 parameters describing the halo model for the star formation history, described in Section \ref{sssec:methods.theo.cib_sfr}:  $\{m_0,m_z,\eta_0,\eta_z\}$.
    We also marginalise over two intrinsic alignment parameters $A_0^{\rm DES/KiDS}$, describing the amplitude of IAs in the DES and KiDS samples. For both parameters we use a Gaussian prior centered at $A_0=1$ with standard deviation $\sigma(A_0)=0.5$. This choice encompasses the values favoured in the cosmological analyses of \citep{1708.01538,2007.15632}. We fix the IA evolution parameter to $\lambda=0$ in both cases. Finally, we marginalise over one redshift shift parameter $\Delta z^i$ for each redshift bin (5 for KiDS, 4 for DES). For these, we use the Gaussian priors described in \citep{1708.01538,2007.15632}. Shape measurement uncertainties were propagated by marginalising over one multiplicative bias parameter per redshift bin. This was done analytically, following the procedure of \cite{2021A&A...646A.129J} with the Gaussian priors of \citep{1708.01538,2007.15632}. In the most general case, our model therefore has 15 free parameters. The priors used for all of them are summarised in Table \ref{tab:priors}.

    We sample this posterior distribution using the Metropolis-Hastings Markov-Chain Monte-Carlo (MCMC) algorithm implemented in \cobaya \citep{2005.05290,2019ascl.soft10019T}. All cosmological theory predictions were computed using the Core Cosmology Library (\ccl, \cite{1812.05995}\footnote{The source code can be found at \url{https://github.com/LSSTDESC/CCL}}). Cosmological parameters were fixed to the \planck{} best-fit values $(\Omega_c,\Omega_b,h,n_s,\sigma_8)=(0.261, 0.049, 0.677, 0.9665, 0.8102)$. We use the halo mass function and halo bias parametrisation of \citet{2010ApJ...724..878T}, with halo masses defined for a spherical overdensity $\Delta=200$ with respect to the critical density.

\section{Data}\label{sec:data}
  \subsection{Cosmic shear}\label{ssec:data.wl}
  \begin{figure}
    \centering
    \includegraphics[width=0.49\textwidth]{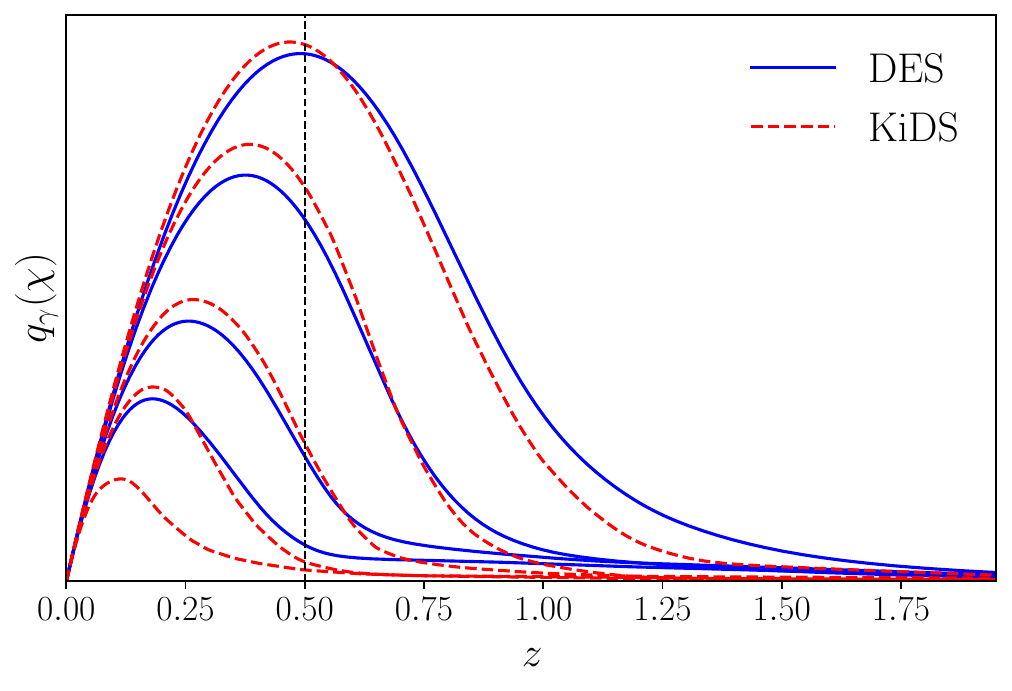}
    \caption{Radial kernels of the 5 \kids (red) and 4 \des (blue) redshift bins used in this analysys. The vertical dashed line marks the redshift at which we evaluate the mean CIB spectra to define the coadding weights.}
    \label{fig:kernels}
  \end{figure}
  \begin{figure*}
    \centering
    \includegraphics[width=0.9\textwidth]{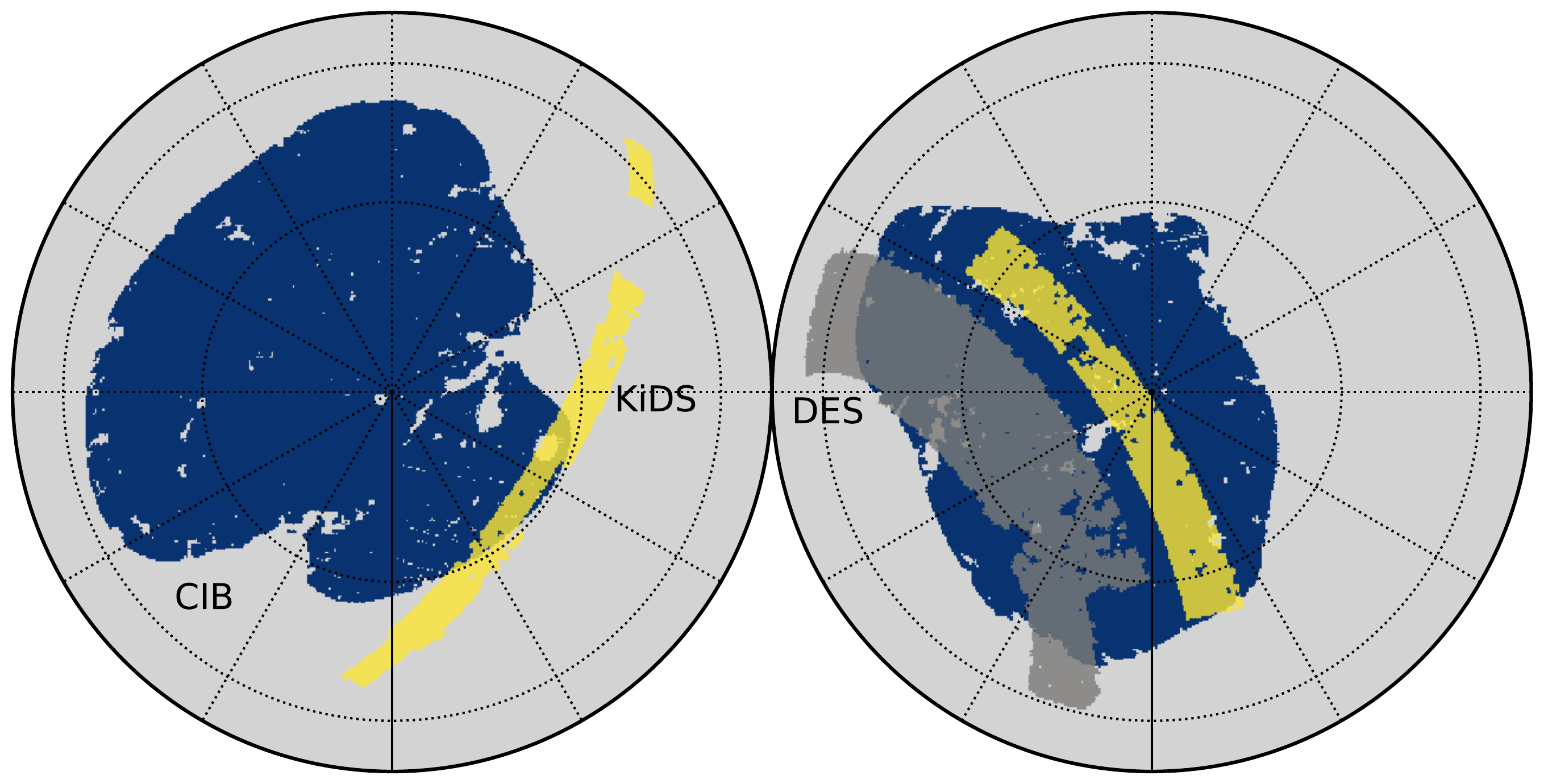}
    \caption{Sky footprint of the CIB and weak lensing surveys used in this analysis in Galactic coordinates.}
    \label{fig:footprint}
  \end{figure*}
    We make use of two weak lensing catalogs released by the Dark Energy Survey and Kilo-Degrey Survey collaborations. The methods used to process these catalogs are described in detail in \cite{2021JCAP...10..030G}, and mostly follow the same procedures used in the official cosmological analyses of both collaborations \citep{1708.01533,1708.01538,2007.01845,2007.15633}. We outline only the main features of these data, and refer the reader to \cite{2021JCAP...10..030G}, as well as the original papers for further details.

    The DES catalog corresponds to the official cosmic shear sample from the first-year (Y1) data release. We make use of the {\tt Metacalibration} catalog, which uses the shape-measurement algorithm~\cite{1702.02601} of the same name to determine individual galaxy ellipticities. The catalog is split into four different tomographic redshift bins spanning the range $z\lesssim1.5$. We make shear maps by averaging the two galaxy ellipticities in each pixel, corrected by the mean response tensor trace. All galaxies are given equal weights, and a mean ellipticity is subtracted from the maps in each redshift bin, following \cite{1708.01538}. We use the official redshift distributions released by DES, marginalising over shifts in the mean redshift as described in Section \ref{sssec:methods.theo.wl}. The catalog covers an area of $\sim1300$ deg$^2$, of which approximately $1000$ deg$^2$ overlap with the the CIB maps used here.

    We also use the cosmic shear catalog publicly released by the KiDS collaboration as part of the fourth data release (DR4), the so-called KiDS-1000 sample \citep{2007.01845}. The sample is divided into the same 5 redshift bins used by the KiDS collaboration, and we use the official redshift distributions released by KiDS to carry out our analysis (again marginalising over mean shifts). Galaxy shear is estimated using {\tt lensfit} \citep{1210.8201}, and we generate shear maps taking the corresponding shape measurement weights into account. We correct for a mean multiplicative bias and subtract the residual mean ellipticity in each redshift bin. The catalog covers an area of $\sim1000$ deg$^2$, of which $\sim500$ deg$^2$ overlap with the CIB footprint.

    Figure \ref{fig:kernels} shows the weak lensing kernels, defined in Eq. \ref{eq:kernel_wl}, corresponding to each of the 9 redshift bins (4 DES bins in solid blue, 5 KiDS bins in dashed red). The four DES bins lead to lensing kernels that are remarkable similar to the four higher-redshift KiDS bins. The weak lensing signal in the first KiDS bin is rather small and, as we will see, contributes negligibly to our cross-correlation analysis. The last two redshift bins, which dominate the cross-correlation signal, peak at around $z\sim0.5$ (marked with a dashed black vertical line in the figure), and thus our measurements are mostly sensitive to the star formation history in this range of redshifts. The sky footprints of both samples are shown in Fig. \ref{fig:footprint}, together with that of the CIB maps described in the next section.

  \subsection{CIB maps}\label{ssec:data.CIBLenz}
    \subsubsection{Multi-frequency maps}\label{sssec:data.CIBLenz.multi}
      For our analysis, we use the CIB maps constructed by \citet{1905.00426} from the \planck{} 353, 545 and 857 GHz temperature maps, corrected for contamination from Galactic dust using neutral hydrogen data from the HI4PI survey \citep{1610.06175}. As in \JRGKA{}, we use the $20\%$ sky masks provided with these data, with an ${\rm HI}$ column density threshold of $N_{\rm HI} > 2.5 \times 10^{20}\,{\rm cm}^{-2}$, apodised with a $15'$ FWHM Gaussian kernel. All power spectra were corrected for the effective beam of each of these maps, which is also provided in the public data release.

      To avoid the loss of power on large scales reported by \cite{1905.00426}, due to the local removal of Galactic dust in overlapping sky patches, we exclude multipoles $\ell<100$ from the analysis.

    \subsubsection{A coadded CIB map}\label{sssec:data.CIBLenz.coadd}
      As shown in \citet{2022arXiv220401649Y,2206.15394}, the cross-correlation signal of the three different frequency maps is highly ($90-95\%$) correlated. This has the advantage that the analysis can be carried out on each frequency map independently, and the results can then be compared to validate the internal consistency of the model used to connect the CIB to the SFR history. The disadvantage, however, is that care must be taken when combining power spectra from all frequencies, since a slight mis-estimation of the cross-covariance terms involving different frequency channels can lead to artificially high or low $\chi^2$ values and, in the worst case, to a bias in the final parameter constraints. In \JRGKA{} we showed that the analytical covariance matrix had to be corrected by a few percent in these cross-terms in order to match the jackknife estimate. Such a small correction is normally irrelevant, however in this case it can change the $\chi^2$ values by up to a factor 2, due to the tight correlation between frequencies. Since the current analysis covers a wider range of scales than that of \JRGKA{}, which might require a more careful modelling of this correction, we have opted to avoid it altogether.
    \begin{figure*}
      \centering
      \includegraphics[width=0.7\textwidth]{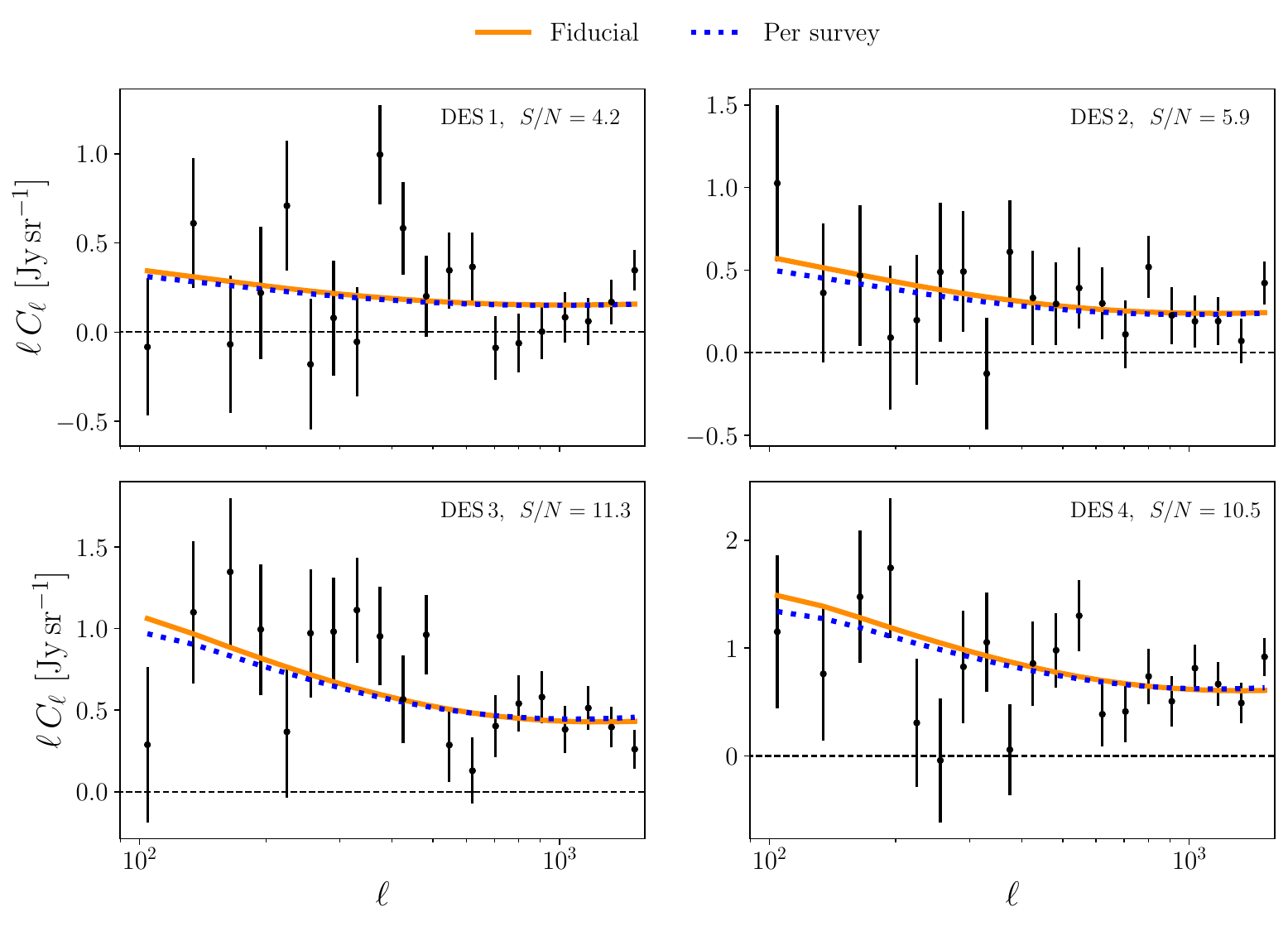}
      \includegraphics[width=0.9\textwidth]{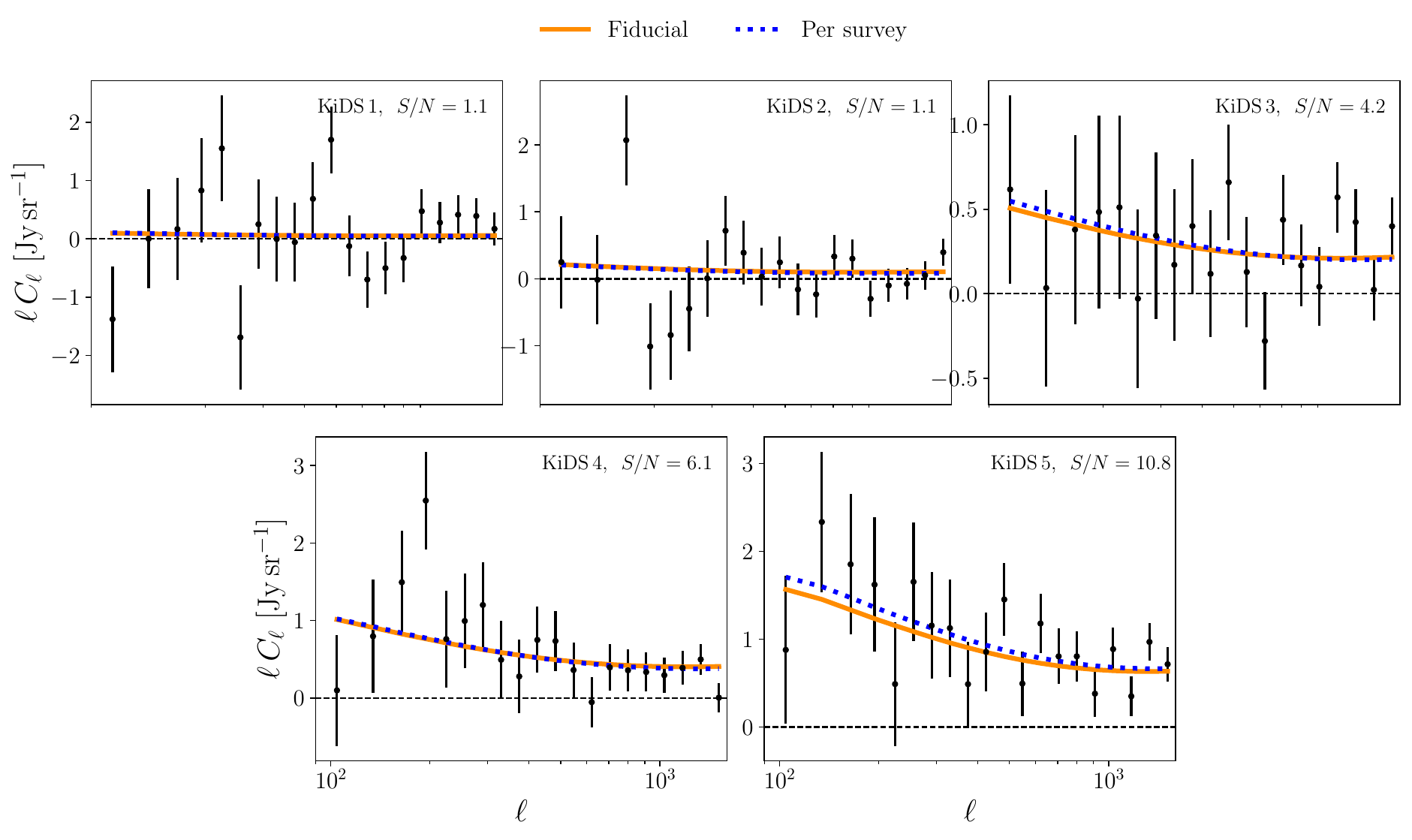}
      \caption{{\sl Top panels:} cross-power spectra between the coadded CIB map and the 4 DES tomographic samples. {\sl Bottom panels:} as above for the 5 KiDS tomographic bins. The orange solid lines show the best-fit model found by fitting all power spectra simultaneously, while blue dotted lines show the best-fit model for each survey (DES or KiDS) independently. We find consistent results between both surveys.}\label{fig:cl_coadd}
    \end{figure*}

      To do so, we construct a single co-added map as an optimal linear combination of the three frequency maps. To determine the linear coefficients of this combination we model the three frequency maps as:
      \begin{equation}\label{eq:cib_lin}
        {\bf m}(\nv)={\bf s}\,c(\nv)+{\bf n}(\nv),
      \end{equation}
      where ${\bf m}(\nv)$ is a 3-element vector containing the three frequency maps at sky position $\nv$, ${\bf s}$ is a vector containing the effective CIB spectrum, $c(\nv)$ is the coadded map we are trying to reconstruct, and ${\bf n}(\nv)$ is the contribution from instrumental noise. Assuming Gaussian noise, an optimal estimator for $c$ can then be found via least-squares minimisation to be
      \begin{equation}\nonumber
        \hat{c}(\nv)={\bf w}^T{\bf m}(\nv),\hspace{12pt}{\bf w}^T=\frac{{\bf s}^T{\sf N}^{-1}}{{\bf s}^T{\sf N}^{-1}{\bf s}},
      \end{equation}
      where ${\sf N}$ is the noise covariance matrix, and ${\bf w}$ are the linear weights we were seeking.

      One must note that Eq. \ref{eq:cib_lin} is only correct in the limit where the CIB frequency maps are 100\% correlated at the signal level. Within the model used here (Eq. \ref{eq:cib_proj}), this would only be strictly valid if the frequency and redshift dependence of the effective infrared spectra ($S_\nu^{\rm eff}(z)$ in Eq. \ref{eq:kernel_cib}) were factorisable. Nevertheless, since correlation between the three maps is very high, assuming a perfect correlation in order to find ${\bf w}$ will only lead to a small loss of sensitivity when using the resulting coadded map instead of all the frequency maps. Since, as we have shown, the weak lensing kernel peaks in the range $z\sim0.5$ for the cosmic shear maps, we construct the vector of spectra ${\bf s}$ by evaluating $S_\nu^{\rm eff}$ at a pivot redshift $z_0=0.5$. In detail, in order to preserve the units of the coadded CIB map (${\rm MJy}/{\rm sr}$) we construct the elements of ${\bf s}$ as:
      \begin{equation}
        s_\nu = S^{\rm eff}_\nu(z_0)/S^{\rm eff}_{857}(z_0),
      \end{equation}
      where $S^{\rm eff}_{857}$ is the spectrum in the 857 GHz channel (our highest signal-to-noise map). The coadded map will therefore have an amplitude comparable to that of the 857 map.
    
      To estimate the noise covariance matrix ${\sf N}$ we first compute the noise power spectrum of the frequency maps $N^{\nu\nu'}_\ell$ (including all auto- and cross-correlations) using the difference between half-mission maps to null out the signal. Each entry of ${\sf N}$ is then calculated by averaging $(2\ell+1)N^{\nu\nu'}_\ell$ in the range $100\leq\ell\leq1500$ used here. The resulting set of coadd weights is:
      \begin{equation}
        (w_{353},w_{545},w_{857})=(0.06,0.22,0.92).
      \end{equation}
      We found that varying the pivot redshift in the range $0.2\leq z_0\leq1$, as well as the range of multipoles used to determine ${\sf N}$ did not change the results significantly.

      Finally, since the different frequency maps have different masks and effective beams, in practice
      we coadd the frequency maps at the level of the power spectra. The resulting cross-power spectrum of cosmic shear with the coadded map, and its covariance become:
      \begin{align}
        &C^{\gamma,{\rm co}}_\ell=\sum_\nu w_\nu\,C^{\gamma,\nu}_\ell,\\
        &{\rm Cov}(C^{\gamma,{\rm co}}_\ell,C^{\gamma',{\rm co}}_{\ell'})=\sum_{\nu\nu'} w_\nu w_{\nu'}\,{\rm Cov}(C^{\gamma,\nu}_\ell,C^{\gamma',\nu'}_{\ell'}).
      \end{align}
      Since the weights are scale-independent, this is equivalent to a linear combination at the map level.

\section{Results}\label{sec:results}
  \subsection{Power spectra, validation, and goodness of fit}\label{ssec:results.cls}
    \begin{figure}
      \centering
      \includegraphics[width=0.48\textwidth]{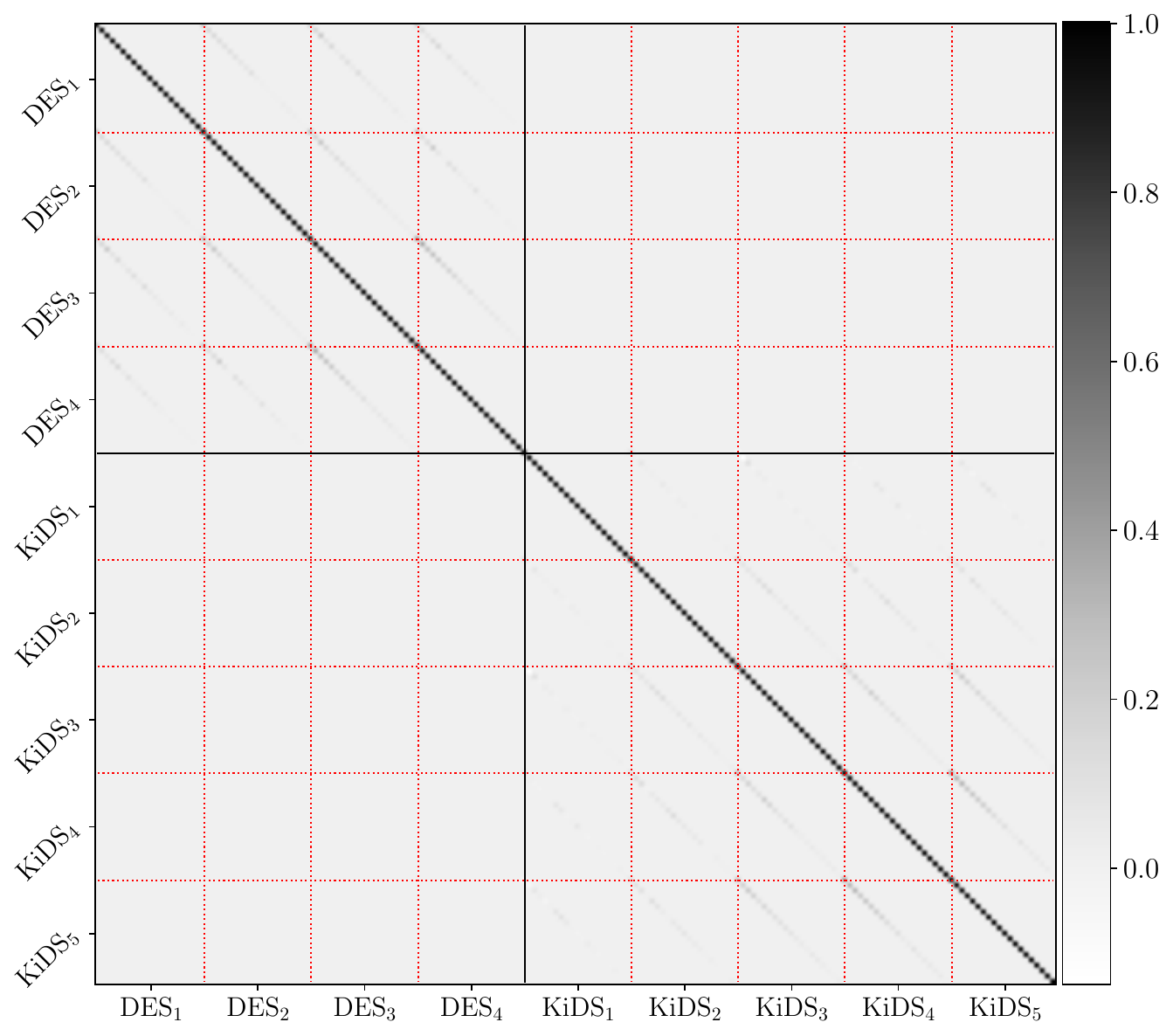}
      \caption{Correlation matrix $r_{ij}$ of the total data vector. The first $4\times4$ and  the last $5\times5$ blocks correspond to the cross-correlations with the coadded CIB map of the DES and KiDS tomographic bins respectively. The off-diagonal elements between DES and KiDS $C_\ell$s are zero, since both surveys do not overlap on the sky.} \label{fig:covmat}
    \end{figure} %
    \begin{figure*}
      \centering
      \includegraphics[width=0.9\textwidth]{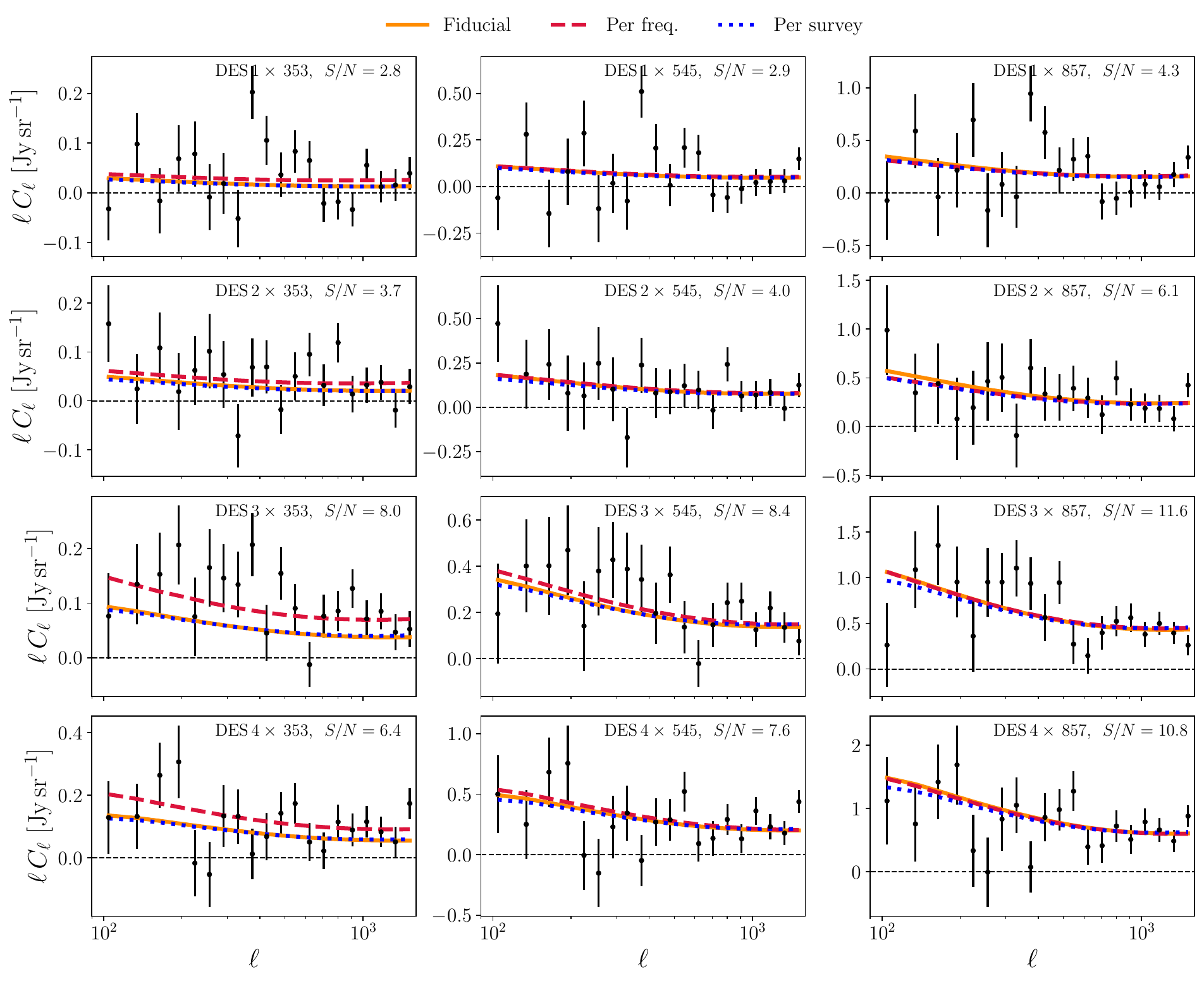}
      \caption{Cross-power spectra between the four DES tomographic bins (from top to bottom) and the 3 CIB frequency maps (353, 545, and 857 GHz from left to right). The points with error bars show our measurements. As in Fig. \ref{fig:cl_coadd}, the orange solid line shows the best-fit model prediction obtained by analysing all the cross-spectra with the coadded CIB map, while the dotted blue line shows the best fit obtained using only the cross-correlations with the DES data. The red dashed lines show the best-fit model obtained from the cross-correlations of all tomographic bins with each frequency map separately. The best-fit model obtained from the coadded map marginally under-estimates the correlation although, as described in Section \ref{ssec:results.robust}, the resulting model constraints are compatible within the statistical uncertainties. Similar results are found for the KiDS data, which we do not show here for brevity.}\label{fig:cl_DES_freq}
    \end{figure*}
    The data vector that will form the basis of our fiducial analysis is the set of 9 cross-correlations between the coadded CIB map described in Section \ref{sssec:data.CIBLenz.coadd} and the cosmic shear $E$-mode in the 4 DES and 5 KiDS tomographic redshift bins. These measurements and their statistical uncertainties are shown in Fig. \ref{fig:cl_coadd}. We obtain clear detections of the cross-correlation especially at the higher redshifts, where the amplitude of both the cosmic shear and CIB signals grows. As a first model-independent estimate of the total signal-to-noise ratio ($S/N$) of this cross-correlation, we find
    \begin{equation}\label{eq:sn1}
      S/N=\sqrt{\chi^2_0-N_d}=20.3,
    \end{equation}
    where $\chi_0^2$ is the $\chi^2$ statistic for a null hypothesis (i.e. $\chi_0^2\equiv {\bf d}^T{\sf C}_{\bf d}^{-1}{\bf d}$ in the notation of Section \ref{ssec:methods.like}), and $N_d=180$ is the size of the data vector. This is, therefore one of the highest-significance detections of this cross-correlation (see also \cite{2109.04458}).

    The orange solid lines in Fig. \ref{fig:cl_coadd} show the best-fit theory prediction for the SFR model of \MRG{} (see Section \ref{sssec:methods.theo.cib_sfr}), found by maximising the posterior distribution of the full set of 9 power spectra. The dashed blue lines, in turn, show the best-fit model obtain by fitting the power spectra corresponding to DES and KiDS separately. The similarity between both best-fit models in all cases, compared with the size of the error bars, therefore implies that the measurements made with both surveys are consistent (we will show this over the full parameter space in the next section). The overall best-fit model has a $\chi^2=187.6$, which corresponds to an acceptable probability-to-exceed (PTE) of $p=0.28$ assuming $N_d-4=176$ degrees of freedom\footnote{We subtract only the 4 SFR parameters from the total number of datapoints when computing the number of degrees of freedom, since all other model parameters have tight priors. In any case, subtracting all model parameters yields a lower, but still acceptable probability $p=0.12$.}. Having a best-fit model at hand allows us to compute the detection significance in a different way, as
    \begin{equation}
      S/N=\sqrt{\chi^2_0-\chi^2_{\rm BF}}=20.1,
    \end{equation}
    where $\chi^2_{\rm BF}$ is the $\chi^2$ of the best-fit model, and $\chi_0^2$ was defined after Eq. \ref{eq:sn1}. The different panels in Fig. \ref{fig:cl_coadd} include the signal-to-noise of each cross-correlation calculated in this manner. We detect the cross-correlation above $4\sigma$ significance in all the DES redshift bins and in the 3 highest redshift KiDS bins, whereas the two lowest redshift KiDS bin contribute negligibly to the signal. The difference in the detectability of the signal in the first DES bin and the second KiDS bin (which have similar kernels, as shown in Fig. \ref{fig:kernels}) is due to the higher area overlap of the DES footprint.

    Since weak lensing should lead to an undetectable $B$-mode signal (at least given current sensitivities), ensuring that the measured $B$-mode power spectra are compatible with zero is a good test for potential systematics in the data. As shown in Appendix \ref{app:bmodes}, we find that the total $B$-mode signal is compatible with zero, and that this is also the case for most of the individual power spectra. However, we find that the cross-correlation of the second KiDS bin with the coadded CIB map yields an unacceptably low $\chi^2$ probability ($p=0.001$) for a null signal. We find this to be the case for the cross-correlation of this bin with the individual CIB frequency maps as well. This may be a statistical fluke, or a sign of systematics in this sample, which was also flagged specifically in the official cosmic shear analysis by the KiDS collaboration \citep{2007.15633}. Since this particular bin does not contribute significantly to the model constraints obtained here, we have included it in our analysis, but verified that our results are unaffected by this choice (see Section \ref{ssec:results.robust}).

    Fig. \ref{fig:covmat} shows the correlation matrix of the power spectra shown in Fig. \ref{fig:cl_coadd}. There are significant off-diagonal correlations (at the level of $30-40\%$) between different redshift bins in the same survey, which extend to almost all pairs of bins due to the cumulative nature of weak lensing. The covariance between the DES and KiDS measurements is zero since their footprints do not overlap.

    We have also studied the cross-correlations with the individual CIB frequency maps. The $12$ cross-correlations between the three frequency maps and the 4 DES redshift bins are shown in Figure \ref{fig:cl_DES_freq}, with qualitatively equivalent results found for KiDS. As before, the solid orange lines show the best-fit model found using all the cross-correlations with the coadded CIB map, and the dotted blue line shows the best-fits using only cross-correlations with DES. In turn, the red dashed lines show the best-fit prediction using only the cross-correlations with one of the 3 frequency maps. The fiducial best-fit model is largely consistent with the predictions from the 545 and 857 GHz channels. It, however, consistently under-predicts the 353 GHz power spectra in both DES and KiDS. A similar trend, where the amplitude of the CIB signal is somewhat consistently higher in the 353 GHz channel was also found in cross-correlation with galaxies by \JRGKA{}. This could be due to both systematics in the data (e.g. contamination from CIB or other extra-galactic foregrounds), or a mis-modeling of the CIB spectrum in this channel. Nevertheless, as we will see in Section \ref{ssec:results.robust}, the 353 GHz data lead to constraints on the model parameters that are largely compatible with those of the other two frequencies, and with our fiducial constraints. Thus, and since the coadded CIB map is dominated by the 857 and 545 GHz channels, we have kept the 353 GHz data in our fiducial analysis. Furthermore, we find that the best-fit model from the coadded CIB map is a good fit to the full multi-frequency data vector, with a $\chi^2=547.6$ for $536$ degrees of freedom ($p=0.35$). Calculating the significance of the detection for the multi-frequency data, as described above in the case of the coadded power spectra, yields
    \begin{equation}
      S/N=\sqrt{951.3-547.6}=20.1\,.
    \end{equation}
    This is in striking agreement with the detection significance using only the coadded CIB map, which demonstrates that virtually no information is lost by coadding the 3 CIB maps, due to the tight correlation between them. This has the advantage of simplifying the analysis significantly, since each likelihood evaluation is 3 times faster and, more importantly, we are not sensitive to the impact on the final parameter constraints of a potential mis-modelling of the cross-frequency covariance matrix.

  \subsection{Constraints on star formation models}\label{ssec:results.sfr}
    \begin{figure*}
        \centering
        \includegraphics[width=0.49\textwidth]{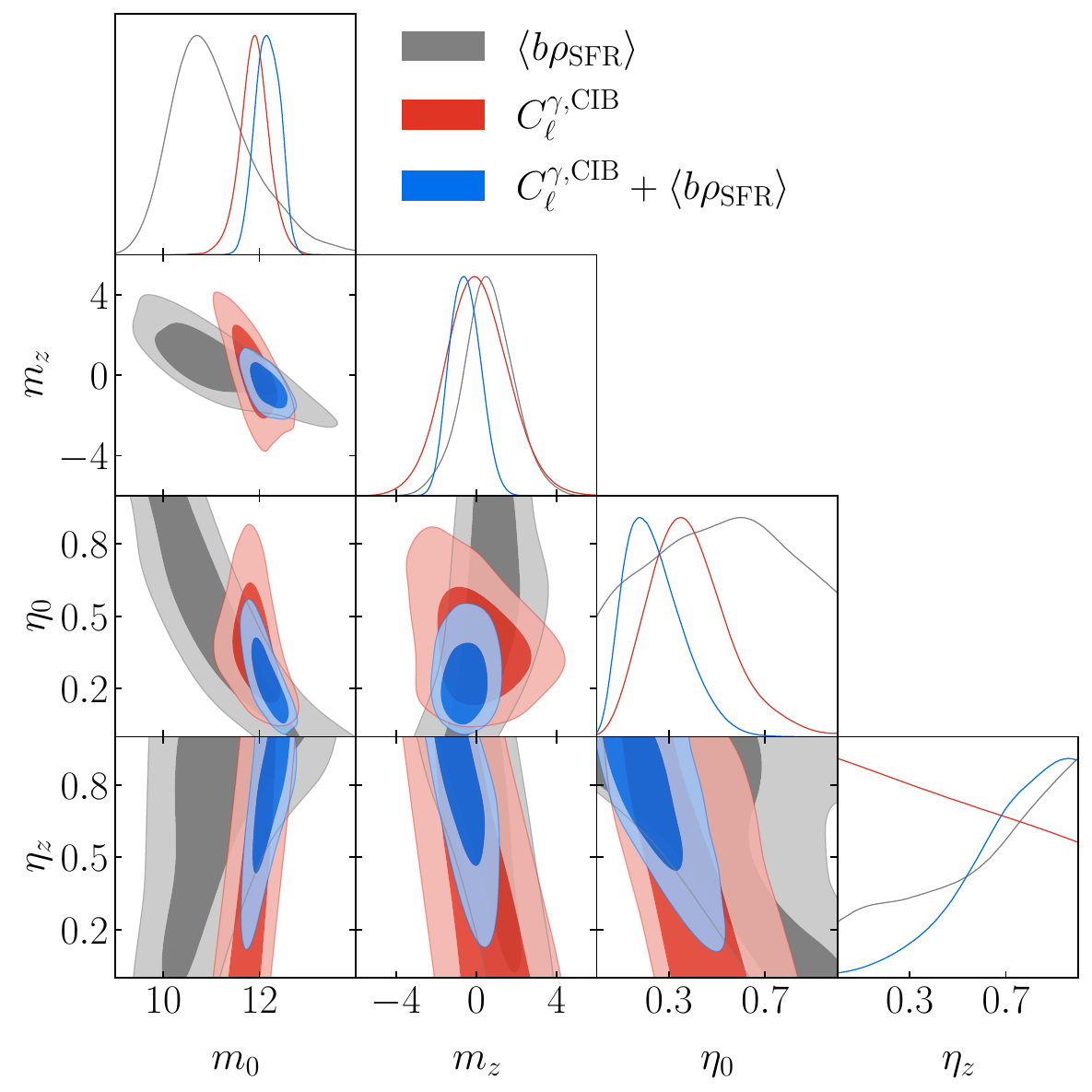}
        \includegraphics[width=0.49\textwidth]{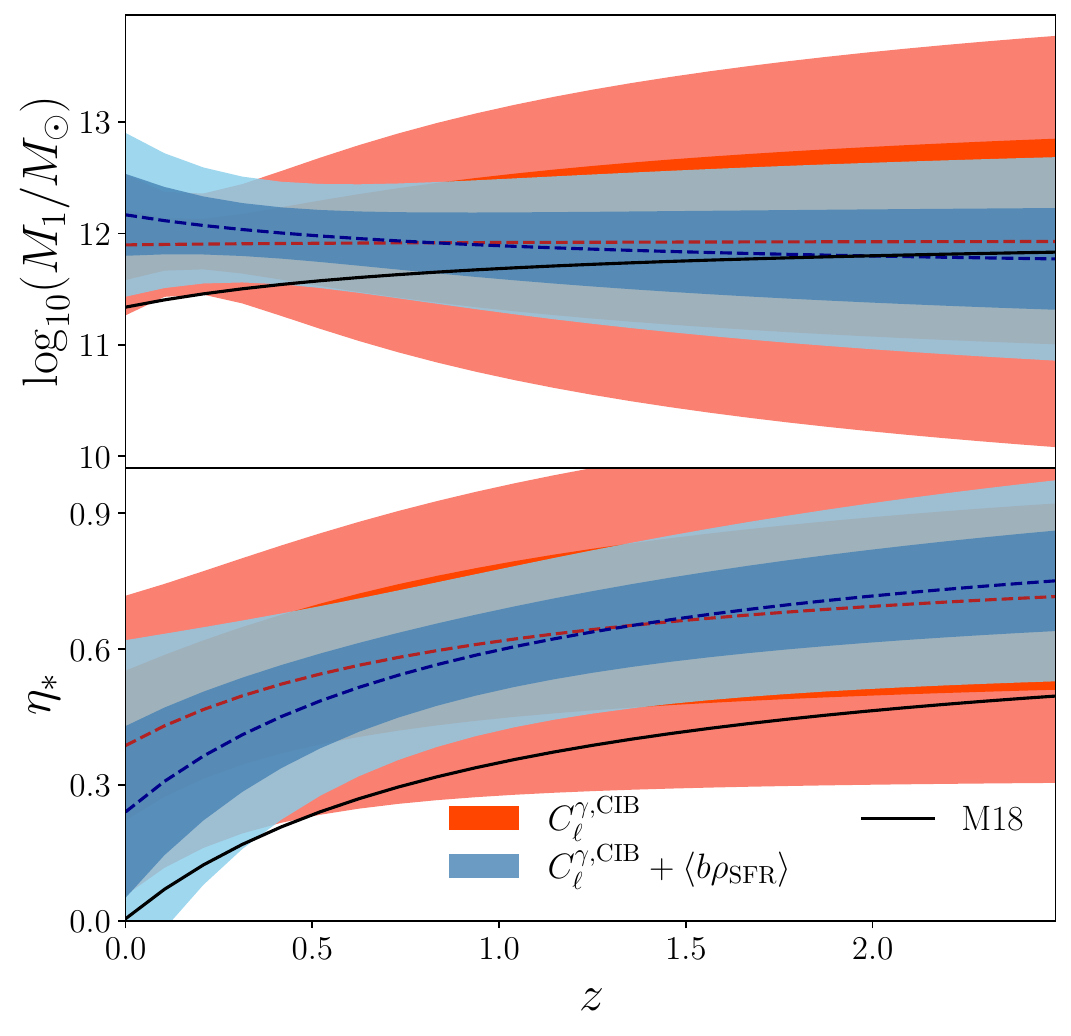}
        \caption{{\sl Left:} constraints on the parameters of the \MRG{} halo-based SFR model, described in Section \ref{sssec:methods.theo.cib_sfr}. The parameter $\eta_0$ describes the maximum efficiency at which gas is transformed into stars at low redshifts, while $\eta_z$ parametrises its redshift evolution. In turn, $m_0$ and $m_z$ parametrise the low-redshift value and evolution of the logarithmic halo mass at which this peak efficiency is attained. Results are shown for our fiducial measurements of the cross-correlation between CIB and cosmic shear data (red), and for its combination with the measurements of $\bsfr$ made by \JRGKA{} (blue). The constraints from the $\bsfr$ data alone are shown in gray. We find that, while we are able to constrain $m_0$ and $m_z$, we can only accurately constrain a linear combination of the two efficiency parameters, corresponding to the value of the peak efficiency at $z\simeq0.4$. {\sl Right:} corresponding constraints on the evolution of the peak efficiency and corresponding peak mass (lower and upper panels) from the shear cross-correlation data (red), and from their combination with the $\bsfr$ measurements. The solid black line shows the best-fit model of \MRG{}.}
        \label{fig:constraints}
    \end{figure*}
  \begin{table}
    \centering
    \def\arraystretch{1.4}
    \begin{tabular}{|llll|}
      \hline
      Data & $m_0$ & $m_z$ & $\eta_{0.4}$\\
      \hline
      $C_\ell^{\gamma,{\rm CIB}}$ & $ 11.90\pm 0.32$ & $ 0.0\pm 1.6$ & $ 0.519^{+0.087}_{-0.18}$\\
      $C_\ell^{\gamma,{\rm CIB}}+\bsfr$ & $ 12.17\pm 0.25$ & $ -0.55^{+0.66}_{-0.87}$ & $ 0.445^{+0.055}_{-0.11}$\\
      $C_\ell^{\gamma,{\rm CIB}}$, DES & $ 12.19\pm 0.39$ & $ -1.5^{+1.9}_{-2.3}$ & $ 0.53^{+0.12}_{-0.26}$\\
      $C_\ell^{\gamma,{\rm CIB}}$, KiDS & $ 11.25^{+0.59}_{-0.44}$ & $ 1.9\pm 2.1$ & $ 0.65^{+0.16}_{-0.25}$\\
      $C_\ell^{\gamma,{\rm CIB}}$, KiDS (no bin 2) & $ 11.31^{+0.66}_{-0.50}$ & $ 1.5^{+2.7}_{-2.3}$ & $ 0.66^{+0.17}_{-0.25}$\\
      $C_\ell^{\gamma,{\rm CIB}}$, 353 GHz & $ 12.27^{+0.37}_{-0.31}$ & $ -1.2^{+1.6}_{-2.0}$ & $ 0.80\pm 0.20$\\
      $C_\ell^{\gamma,{\rm CIB}}$, 545 GHz & $ 11.90^{+0.42}_{-0.33}$ & $ -0.3^{+1.8}_{-2.3}$ & $ 0.64^{+0.15}_{-0.27}$\\
      $C_\ell^{\gamma,{\rm CIB}}$, 857 GHz & $ 11.88\pm 0.31$ & $ 0.1\pm 1.6$ & $ 0.515^{+0.088}_{-0.17}$\\
      $C_\ell^{\gamma,{\rm CIB}}$, no IA & $ 11.62\pm 0.42$ & $ 1.7\pm 1.7$ & $ 0.398^{+0.057}_{-0.10}$\\
      \hline
    \end{tabular}
    \caption{Constraints on the SFR halo model parameters, characterising the mass and redshift dependence of the efficiency with which gas is transformed into star (see Section \ref{sssec:methods.theo.cib_sfr}), for various analysis choices.}\label{tab:results}
  \end{table} %
    Combining the 9 cross-power spectra with the coadded CIB map, we are able to place constraints on the four free parameters of the halo-based SFR model of Section \ref{sssec:methods.theo.cib_sfr}, $\{m_0,m_z,\eta_0,\eta_z\}$. The resulting 1 and 2-$\sigma$ contours are shown in red in the left panel of Fig. \ref{fig:constraints}. The data are able to constrain both the current value of the peak mass, $m_0$, as well as its evolution with redshift $m_z$. In turn, the two parameters describing the redshift evolution of the peak efficiency $\eta_*$, $\{\eta_0,\eta_z\}$, are significantly degenerate and, while we're able to constrain $\eta_0$, $\eta_z$ is completely unconstrained. In particular the data are most sensitive to a linear combination of both parameters that is approximately equal to the value of $\eta_*(z)$ at the peak of the lensing kernel $z=0.4$ of the galaxy redshift bins used in the analysis (see Fig. \ref{fig:kernels}). We will label this parameter $\eta_{0.4}$
    \begin{equation}
      \eta_{0.4}\equiv\eta_0+\eta_z\frac{0.4}{1+0.4}.
    \end{equation}
    The constraints on these three parameters, $\{m_0,m_z,\eta_{0.4}\}$, from the shear-CIB power spectra, are listed in the first row of Table \ref{tab:results}. The right panel in Fig. \ref{fig:constraints} shows, in red, the 1 and 2$\sigma$ bounds on the evolution of the two SFR functions, $M_1$ (top) and $\eta_*$ (bottom), derived from our constraints on these three parameters. The figure also shows the best-fit model of \MRG{} as a black solid line. Overall, our constraints on $M_1$ are in good agreement with \MRG{}, and recover the predicted $\eta_*$ within $\sim2\sigma$. The data, however, seem to prefer a consistently higher peak star formation efficiency, particularly at low redshifts.

\begin{figure}
        \centering
        \includegraphics[width=0.49\textwidth]{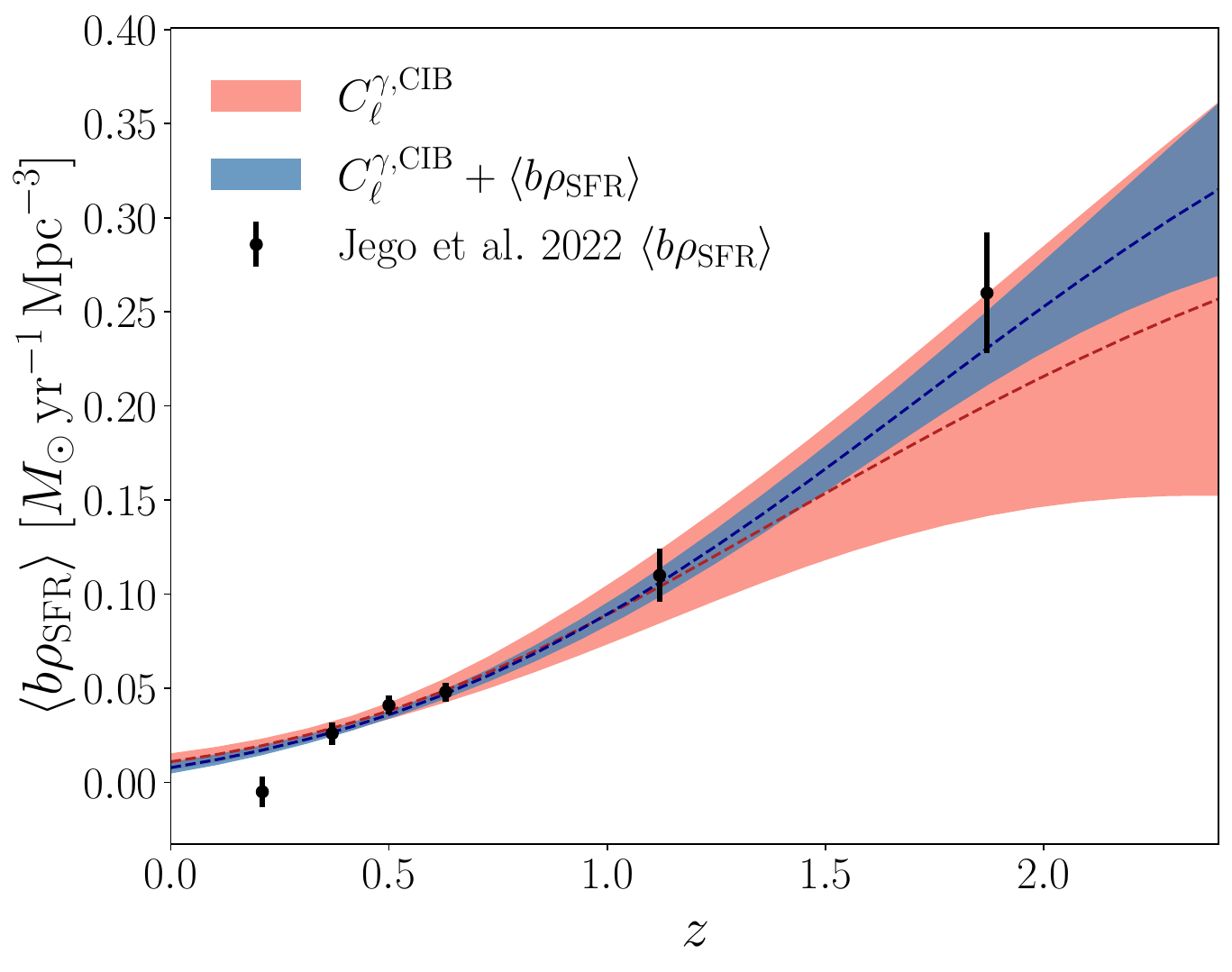}
        \caption{Bias-weighted SFR density as a function of redshift. The measurements of \JRGKA{} are shown as black dots with error bars. The $1\sigma$ constraints derived from our measurements of the shear-CIB cross-correlation are shown as a shaded red band, while the constraints combining both datasets are shown in blue.}
        \label{fig:bsfr}
    \end{figure}

    It is interesting to explore to what extent our constraints are in agreement with other measurements of the star formation history from CIB data. In particular, recently \JRGKA{} made model-independent measurements of the ``bias-weighted SFR density'' $\bsfr$ (see Eq. \ref{eq:bsfr}) through cross-correlations of the CIB with galaxy samples in the range $0\lesssim z\lesssim2$. Fig. \ref{fig:bsfr} shows the measurements of $\bsfr$ of \JRGKA{} (black points) together with the 1$\sigma$ confidence interval on this quantity derived from our constraints on the SFR model parameters (red band). Our constraints are in reasonable agreement with the $\bsfr$ measurements, with a marginal preference for lower values at high redshifts.
    \begin{figure*}
        \centering
        \includegraphics[width=0.99\textwidth]{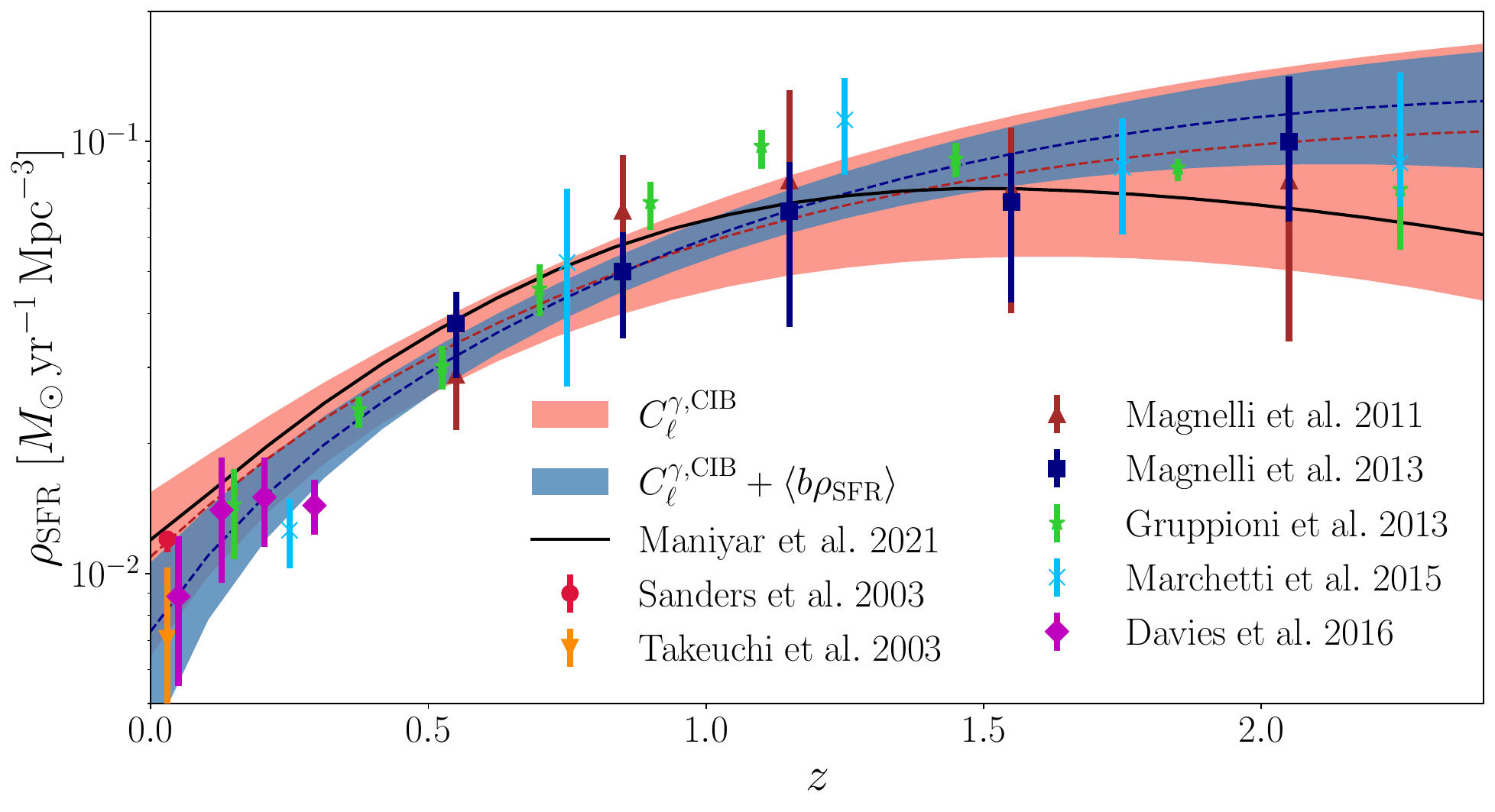}
        \caption{Star formation rate density as a function of redshift. The points with error bars correspond to the direct measurements from the infrared luminosity function by various authors (see legend and main text). The solid black line shows the best-fit prediction from the CIB analysis of \citet{2021A&A...645A..40M}. Our constraints from the cross-correlation between the CIB and cosmic shear data is shown as a red shaded band. Adding to these the  measurements of the bias-weighted SFR density $\bsfr$ of \JRGKA{} yields the blue shaded band (both bands show the 68\% confidence intervals).}
        \label{fig:rsfr}
    \end{figure*}

    Given the compatibility of our measurements of $C_\ell^{\gamma,{\rm CIB}}$ with the $\bsfr$ data, we can improve our constraints on the SFR model parameters by combining both datasets. This can be done easily by importance-sampling the MCMC chains obtained from the shear power spectrum measurements. Since, as shown in \JRGKA{}, the uncertainties on the measured $\bsfr$ are Gaussianly distributed, we assign a multiplicative weight to each sample $i$ given by
    \begin{equation}
      w_i\propto\exp\left[-\frac{1}{2}(\hat{\bf b}-{\bf b}_i)^T{\sf C}_{\bf b}^{-1}(\hat{\bf b}-{\bf b}_i)\right].
    \end{equation}
    Here $\hat{\bf b}$ is a vector with the six measurements of $\bsfr$, ${\sf C}^{-1}_{\bf b}$ is the covariance matrix of those measurements (also provided in \JRGKA{}), and ${\bf b}_i$ is the prediction for $\bsfr$ using the the model parameters of the $i$-th sample. It is worth noting that this procedure assumes that the $\bsfr$ measurements are uncorrelated with the angular power spectra used in this analysis. This is not entirely true. At $z\lesssim0.8$, the $\bsfr$ data were measured using data from the DESI Legacy Survey \citep[DELS]{1804.08657}, which partially overlaps with the \kids sample used here. However, the $\bsfr$ measurements were carried out using the full area overlap between DELS and the CIB maps used here, of which only a small fraction coincides with the \kids footprint. Furthermore, the uncertainties in the power spectra measured here receive a significant contribution from shape noise in the \kids sample, which is uncorrelated with the DELS data. Thus, the actual correlation between the $\bsfr$ measurements and the cross-correlation data used here is minimal. We estimate that the correlation between the CIB-shear power spectra used here, and the DELS$\times$CIB cross-correlations used in \cite{2206.15394} is below 10$\%$ for all scales used here. We therefore neglect this correlation in our analysis.
    
    The gray contours in the left panel of Fig. \ref{fig:constraints} show the constraints on the SFR parameters obtained solely from the $\bsfr$ measurements (which coincide with the results published by \JRGKA{}). The blue contours, in turn, show the constraints obtained from the combination of the shear-CIB $C_\ell$s and the $\bsfr$ measurements. These are reported in the second row of Table \ref{tab:results}. The constraints on $m_0$, $m_z$, and $\eta_{0.4}$ improve by $\sim22\%,\,\,52\%$, and $43\%$ respectively after including the $\bsfr$ data. The corresponding constraints on the evolution of $M_1$ and $\eta_*$ from this combination are shown in blue in the right panel of Fig. \ref{fig:constraints}. Although $M_1$ is still in good agreement with the best-fit model of \MRG{} (black solid line), adding the $\bsfr$ data accentuates the preference for a higher $\eta_*$, particularly at high redshifts. This, however, is mostly driven by the $\bsfr$ data, since the cosmic shear data used here is not very sensitive to redshifts $z\gtrsim1.0$. The resulting constraints on $\bsfr$ are shown in blue in Fig. \ref{fig:bsfr}. Unsurprisingly, the results are in good agreement with the direct measurements of \JRGKA{}.

    Finally, we compare the evolution of the star formation rate density predicted from our model constraints with independent direct measurements of $\rsfr$ based on measurements of the infrared luminosity function. The results are shown in Fig. \ref{fig:rsfr}, where the different points with error bars show the measurements of \citet{astro-ph/0306263,astro-ph/0303181,1101.2467,2013A&A...553A.132M,2013MNRAS.432...23G,2016MNRAS.456.1999M,2016MNRAS.461..458D}, some collected by \citet{2014ARA&A..52..415M}. The red band shows the 68\% constraints derived from our measurements of the CIB-weak lensing cross-correlation, while the blue band shows the combination with the $\bsfr$ measurements. Our constraints are generally in good agreement with all independent measurements of the SFR density covering the last $\sim10$ billion years of cosmic history. This reinforces the interpretation of the CIB as being sourced primarily by the re-emission of UV light absorbed by dust in star-forming galaxies. The solid black line in the same figure shows the best-fit star formation model found by \citet{2021A&A...645A..40M} from the analysis of the CIB auto-correlation. Although our measurements are broadly compatible with this preferred model, the combination of $C_\ell^{\gamma,{\rm CIB}}$ and $\bsfr$ data predicts a higher SFRD at redshift $\sim2$. As pointed out by \JRGKA{}, this is mostly driven by the high-redshift $\bsfr$ data, based on the large-scale cross-correlation between the CIB and quasars. An alternative tomographic cross-correlation with the CIB at high redshifts would thus be useful to confirm or rule out this trend. This could be achieved through the cross-correlation with maps of the CMB lensing convergence, assuming that the systematics arising from cross-contamination with the CIB can be kept under control.

  \subsection{Robustness tests}\label{ssec:results.robust}
    \begin{figure*}
        \centering
        \includegraphics[width=0.49\textwidth]{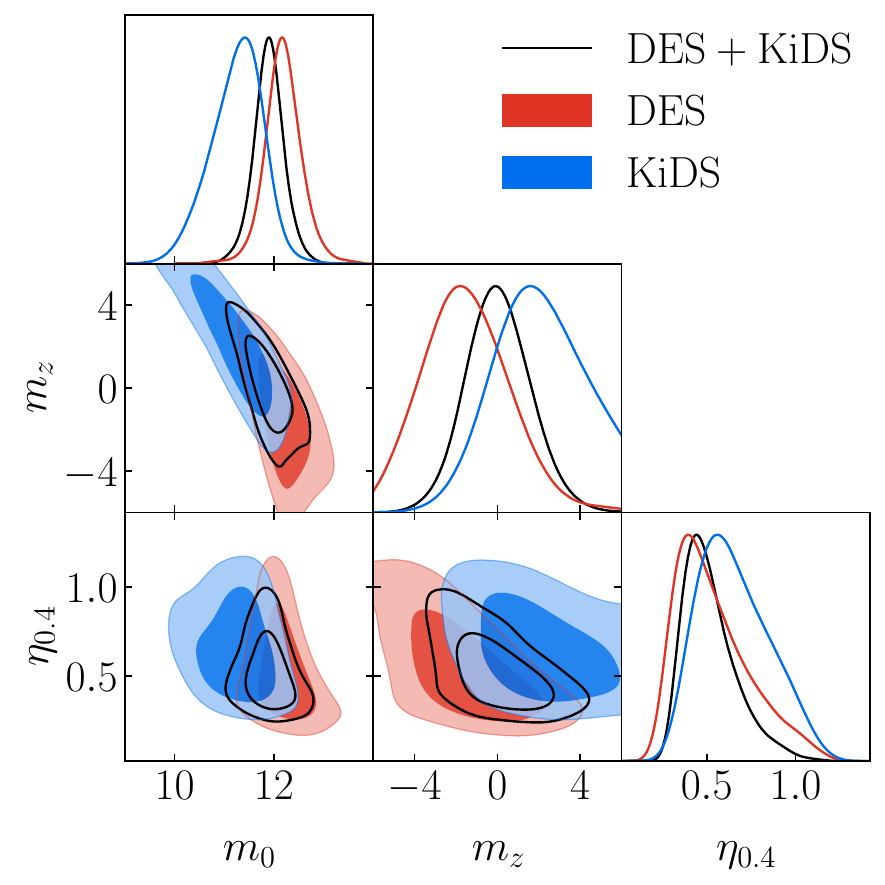}
        \includegraphics[width=0.49\textwidth]{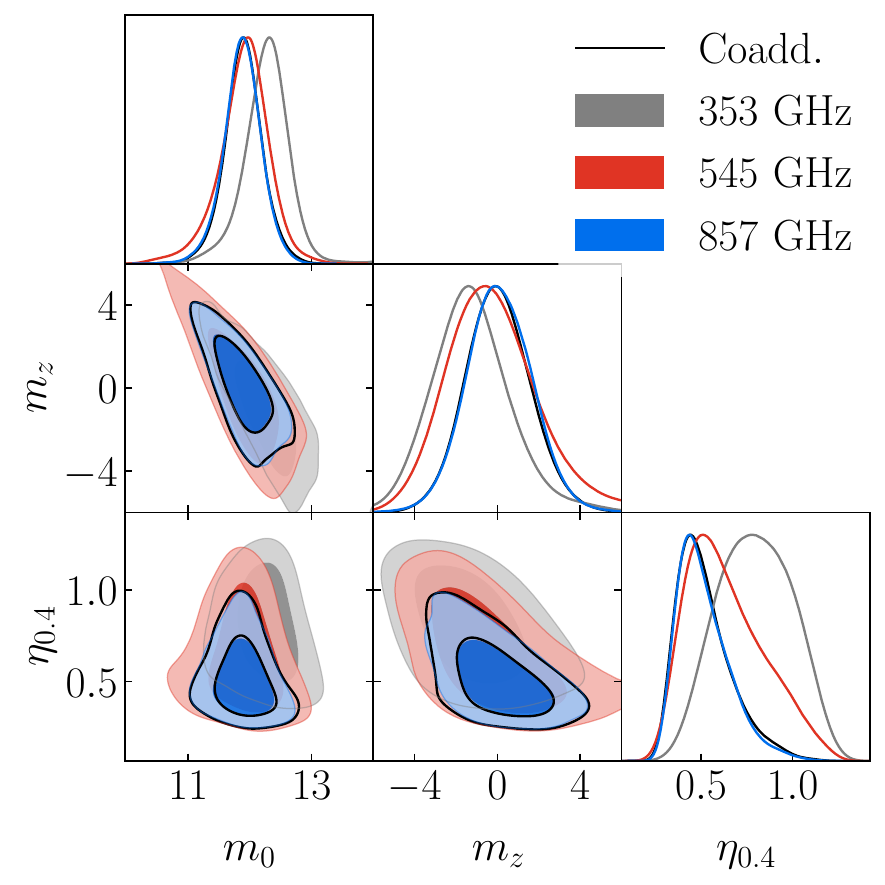}
        \caption{Robustness tests of our main results. The {\sl left panel} shows the constraints on the SFR halo model parameters found using our full data vector of cross-correlations (solid black lines), while the red and blue contours show the results of considering only the cross correlations with DES or KiDS respectively. The {\sl right panel} shows the dependence of the final results on the choice of CIB map. Our fiducial results from the coadded map are shown as solid black lines, while the results using the 353, 545, and 857 GHz maps are shown in gray, red, and blue, respectively.}
        \label{fig:constraints_robust}
    \end{figure*}
    We have quantified the robustness of the results presented in the previous sections to various choices made in our analysis, as well as various potential sources of systematic uncertainty. First, in order to test for the potential impact of systematics in the cosmic shear measurements, which would affect the DES and KiDS datasets differently, we have re-run MCMC chains using only one of the two datasets. The resulting constraints are shown in the left panel of Fig. \ref{fig:constraints_robust} in red and blue for DES and KiDS respectively, together with our fiducial combined constraints as black solid lines. We show results for the three parameters that our data are able to constrain: $\{m_0,m_z,\eta_{0.4}\}$, and the corresponding 1$\sigma$ bounds are listed in Table \ref{tab:results}. The constraints found with either dataset are compatible within $\sim1\sigma$. However, we find that the KiDS dataset has a preference for lower peak efficiency mass values. This is in line with the results from the cosmological analysis of the KiDS collaboration \citep{2007.15633,2007.15632}, where the weak lensing amplitude, parametrised by $S_8\equiv\sigma_8\sqrt{\Omega_m/0.3}$, was found to be lower than the preferred \planck{} value at the $\sim3\sigma$ level. Since we use the best-fit \planck{} cosmology as fiducial in our analysis, a lower halo mass value (and hence a lower halo bias) is required in order to match the KiDS data. Cosmological constraints from weak lensing data are sensitive to systematics in the calibrated redshift distributions used to interpret them. Although we have made use of the official redshift distributions, and marginalised over systematic uncertainty parameters as done by both collaborations, there is some evidence of potential mis-calibration of the DES $p(z)$s that could lower $S_8$ to values that are more in line with the KiDS data \citep{1906.09262}. Regardless of this, the differences between both datasets observed here are small in relation with the statistical uncertainties, and are compatible with a statistical fluctuation. The main conclusion, nevertheless, is that the constraints on SFR model parameters depend on the background cosmological model assumed. A joint cosmology-SFR analysis making use of all auto- and cross-correlations of CIB and cosmic shear data would therefore be able to robustly account for cosmological uncertainties.

    As an additional test for systematics in the data, we have quantified the significance of the correlations of the CIB intensity with the weak lensing $B$-modes. The details are described in Appendix \ref{app:bmodes}. Although we find that the $B$-mode correlations are compatible with zero, the second KiDS redshift bin yields consistently low $p$-values in all its correlations with the different CIB maps. Although this might be a statistical fluke, the KiDS collaboration also pointed out a similar discrepancy for this bin in their cosmic shear analysis \citep{2007.15633}. Since the $E$-mode cross-correlation with this bin is not detected with any significance, and it contributes negligibly to the final constraints, we have included these correlations in our final analysis. However, as an a posteriori test, we have re-derived our constraints after removing these from the data vector. As shown in the fifth row of Table \ref{tab:results}, removing these data leads to shift in the final parameter values of $\sim0.5\sigma$ or less compared to the constraints used with the full KiDS complement (but without DES). Since our constraints are in fact driven by the DES data, given its larger overlap area, we conclude that including this bin in the final analysis does not affect our final results significantly.

    Our parametrisation of the CIB intensity in terms of $\rsfr$ requires a good description of the mean infrared spectrum at the relevant frequencies. For this we used the measurements of \citet{2013A&A...557A..66B,2015A&A...573A.113B,2017A&A...607A..89B}. In order test the sensitivity of our results to potential inaccuracies in this model, as well as the impact of contamination in the CIB maps from other astrophysical sources (e.g. Sunyaev-Zel'dovich, radio point sources), we have repeated our analysis replacing the coadded CIB map with one of the three single-frequency maps released by \citet{1905.00426}. The results are shown in the right panel of Fig. \ref{fig:constraints_robust} for the 353, 545, and 857 GHz channels in gray, red, and blue respectively, together with our fiducial constraints, shown as black lines. Since, as described in Section \ref{ssec:data.CIBLenz}, our coadded map receives its main contribution from the 857 GHz map, it is not surprising that the constraints obtained with either of these two maps are very similar. It is reassuring, however, to find that these constraints are also highly compatible with those derived from the 545 or 353 GHz maps alone, albeit with larger error bars. In more detail, the data from the 353 GHz map seem to prefer a model with a larger peak efficiency mass: $m_0=12.27^{+0.37}_{-0.31}$ as opposed to $11.90\pm0.32$ in our fiducial constraints, corresponding to a $\sim1\sigma$ shift. Although this is a small shift, in principle, the strong correlation between the three frequency maps makes the difference more significant. This may be a signature of a mis-calibration of the effective CIB spectra used in our model, contamination from other astrophysical foregrounds in the 353 map, or that a more sophisticated model is needed to link CIB intensity and SFR density (e.g. a mass dependence in the shape of the mean infrared spectrum). Nevertheless, these differences are small compared to the final statistical uncertainties for current data.

    Finally, to quantify the impact of intrinsic alignments, we have rederived constraints on the SFR parameters setting the contribution from IAs to zero. The results are shown in the last row of Table \ref{tab:results}. Ignoring intrinsic alignments leads to a $\sim1\sigma$ downwards shift in both the low-redshift peak efficiency $\eta_{0.4}$, and the corresponding halo mass $m_0$. This is in line with what one would expect: assuming a positive IA amplitude, corresponding to the case where galaxy shapes are aligned with the principal eigenvectors of the tidal field, intrinsic alignments have a negative contribution to the cosmic shear signal. Without accounting for this contribution, the only way to fit the amplitude of the measured correlation with the CIB is to lower the peak halo mass (associated with a lower halo bias), or the peak efficiency itself. Given the sensitivity of current data, the effect is relatively small, but it will become more relevant as better weak lensing data become available. The simple LNLA model used here to describe IAs is therefore good enough for our purposes, but more sophisticated frameworks may be required in the future.

\section{Conclusions}\label{sec:conc}
  In this paper we have studied the cross-correlation between maps of the CIB and tomographic measurements of the matter inhomogeneities from cosmic shear data. This data combination is complementary to the study of the CIB auto-correlation, as well as cross-correlations with galaxies, or with the gravitational lensing of the CMB. Since weak lensing probes directly the matter overdensities, this cross-correlation is able to directly test the connection between matter and star formation rate density, in a way that is independent of the complicated relationship between galaxies and matter that plague the cross-correlation with galaxies on small scales. Unlike in the case of CMB lensing, the availability of galaxy redshifts allows cosmic shear cross-correlations to probe the redshift dependence of the CIB, although the cumulative nature of gravitational lensing allows for less flexibility than galaxy cross-correlations. Finally, since cosmic shear catalogs are independent of the CMB data that often forms the basis of CMB lensing and CIB maps, this cross-correlation is less sensitive to instrumental systematics in the CMB data, or to cross-contamination from other astrophysical foregrounds. This cross-correlation is therefore a useful and versatile tool to validate physical models of the CIB, its connection with star formation, and its relation to the underlying matter density fluctuations.

  We have presented measurements of the shear-CIB cross-correlation using the CIB maps of \citet{1905.00426}, and cosmic shear catalogs from the first data release of the Dark Energy Survey \citep{1708.01533}, and from the fourth data release of the Kilo-Degree Survey \citep{2007.01845}. The cross-correlation is detected with a significance of $\sim20\sigma$, which constitutes, to our knowledge, the first high-sensitivity measurement of this signal. Furthermore, we have shown that the measurements can be accurately described by a halo-based model that links the CIB intensity and the star formation rate density within the standard $\Lambda$CDM cosmological model, and which is compatible with other studies of the CIB anisotropies.

  Our measurements have allowed us to place constraints on free parameters of the model that describe the efficiency with which gas is converted into stars as a function of halo mass. In particular, we are able to constrain the halo mass at which this efficiency peaks today, as well as its time evolution, and the value of the peak efficiency around $z\sim0.4$, $\eta_{0.4}$. By combining our cross-correlation data with the measurements of the bias-weighted SFR density of \JRGKA{}, we find that the peak efficiency is $\eta_{0.4}=0.445^{+0.055}_{-0.11}$ and that the corresponding peak mass is
  \begin{equation}
    \log_{10}(M_1/M_\odot) = 12.17\pm0.25+\left(-0.55^{+0.66}_{-0.87}\right)\frac{z}{1+z}.
  \end{equation}
  These results, shown in Fig. \ref{fig:constraints}, are in reasonable agreement with the best-fit model of \MRG{}, although with a marginal preference for a higher peak efficiency. Qualitatively, the results agree also with the peak efficiency and corresponding mass found by \cite{2021A&A...645A..40M,2022arXiv220401649Y}, and confirms the overall picture obtained by previous studies, by which about $40\%$ of all infalling gas is transformed into stars in halos with mass $M\sim 10^{12.5}\,M_\odot$. We have shown that these results are robust with respect to the choice of CIB map and cosmic shear catalog, and are therefore unlikely to be significantly affected by systematics in either dataset. Most importantly, as shown in Fig. \ref{fig:rsfr}, using our parameter constraints to predict the star formation rate density as a function of time from $z\sim2$, we find excellent agreement with independent measurements of $\rsfr$ from the luminosity function of infrared galaxies, further confirming the physical origin of the CIB.

  The analysis presented here has made use of a number of assumptions, that should be more thoroughly tested in the future. First, we have ignored the effect of baryonic feedback on the matter power spectrum when predicting the cosmic shear signal. Although this is likely a subdominant effect on the scales $\ell<1500$ used here, and the theoretical uncertainties in the simple SFR model we use are also probably larger, a more thorough characterisation of the impact of baryons would be useful, particularly in the presence of more constraining cosmic shear data. It would also be interesting to explore whether CIB-derived constraints on star formation could indirectly help constrain feedback effects (e.g. by quantifying the diffuse baryon fraction not accreted into stars, or the rate at which massive, short-lived stars are formed). Secondly, our model assumes a very simple scale dependence of the density of satellite galaxies, following the dark matter distribution through the same NFW profile. This is likely not accurate in detail, and may affect future, more sensitive observations using smaller-scale data, deeper in the 1-halo regime. We leave such a study for future work. Thirdly, we have shown that intrinsic alignments can affect our constraints at the $1\sigma$ level. Here we have used a rather simple model to describe IAs, assuming a direct proportionality with the local tidal field. Although this should suffice for the data studied here, future, more sensitive datasets will likely require the use of more sophisticated models. Furthermore, on the CIB side, we have assumed a simple linear relation to link SFR and infrared luminosity \citep{2003PASP..115..763C}. The value of the proportionality constant, and thus the link between the amplitude of CIB anisotropies and SFR density, crucially depends on the form of the initial mass function at large masses. Although this systematic is common to any constraints on the SFR history, and not just those based on the CIB, the accuracy of this assumption should be more thoroughly tested if a precise understanding of the physics of star formation, and its repercussions on other astrophysical processes, is pursued from these data. Finally, our analysis has assumed perfect knowledge of the background cosmological parameters. This is justified, since these parameters are known to better precision than the SFR model parameters constrained here, thanks to observations of the CMB, large-scale structure, and supernovae. However, given the existing tension between CMB and low-redshift in the values of some of these parameters, particularly in the case of the weak lensing amplitude, a joint analysis of the cosmic shear and CIB data, including all auto- and cross-correlations, and targetting both cosmological and SFR parameters, would allow us to account for cosmological model uncertainties in a self-consistent manner. Furthermore, since photometric redshift systematics may play a significant role in the context of these tensions, future studies of the CIB-shear correlation will also require thorough characterisation and marginalisation of redshift distribution uncertainties.

  In spite of these caveats, our work has shown that important insight on the star formation history, and its dependence on halo mass, can be gained by combining CIB and cosmic shear data. This type of analysis has a promising future, with the advent of new ground-based, high-resolution facilities targetting the far infrared, such as CCAT-prime \citep{2018SPIE10700E..1MS} or the Simons Observatory \citep{1808.07445}, and wide and deep optical weak lensing surveys such as the Rubin Observatory Legacy Survey of Space and Time \citep[LSST,][]{2009arXiv0912.0201L}. The growth in sensitivity and area enabled by these experiments, will vastly increase the significance of the CIB-shear signal, and the range of scales over which it can be effectively employed. In combination with other cross-correlations, including galaxy, CMB lensing, and Sunyaev-Zel'dovich data, this will allow us to improve the our understanding of the physical relation between star formation and the matter distribution, while simultaneously throwing light onto some of the astrophysical modelling uncertainties described above.

\section*{Acknowledgements}
  We would like to thank Martin Rey, Aprajita Verma, and Ziang Yan for useful discussions. BJ is supported by the ENS Paris-Saclay. DA is supported by the Science and Technology Facilities Council through an Ernest Rutherford Fellowship, grant reference ST/P004474. CGG is supported by European Research Council Grant No:  693024 and the Beecroft Trust. JRZ is supported by an STFC doctoral studentship. We made extensive use of computational resources at the University of Oxford Department of Physics, funded by the John Fell Oxford University Press Research Fund. 

  We made extensive use of the {\tt numpy} \citep{oliphant2006guide, van2011numpy}, {\tt scipy} \citep{2020SciPy-NMeth}, {\tt astropy} \citep{1307.6212, 1801.02634}, {\tt healpy} \citep{Zonca2019}, {\tt GetDist} \cite{2019arXiv191013970L}, and {\tt matplotlib} \citep{Hunter:2007} python packages.

\section*{Data Availability}
  The code developed for this work as well as the derived datasets produced (power spectra and covariances) are available upon request. The catalogues and maps used were made publicly available by the authors of the relevant papers, as described in the text.

\bibliographystyle{mnras}
\bibliography{main,non_ads}

\appendix

\section{$B$-mode power spectra}\label{app:bmodes}
  \begin{figure*}
    \centering
    \includegraphics[width=0.7\textwidth]{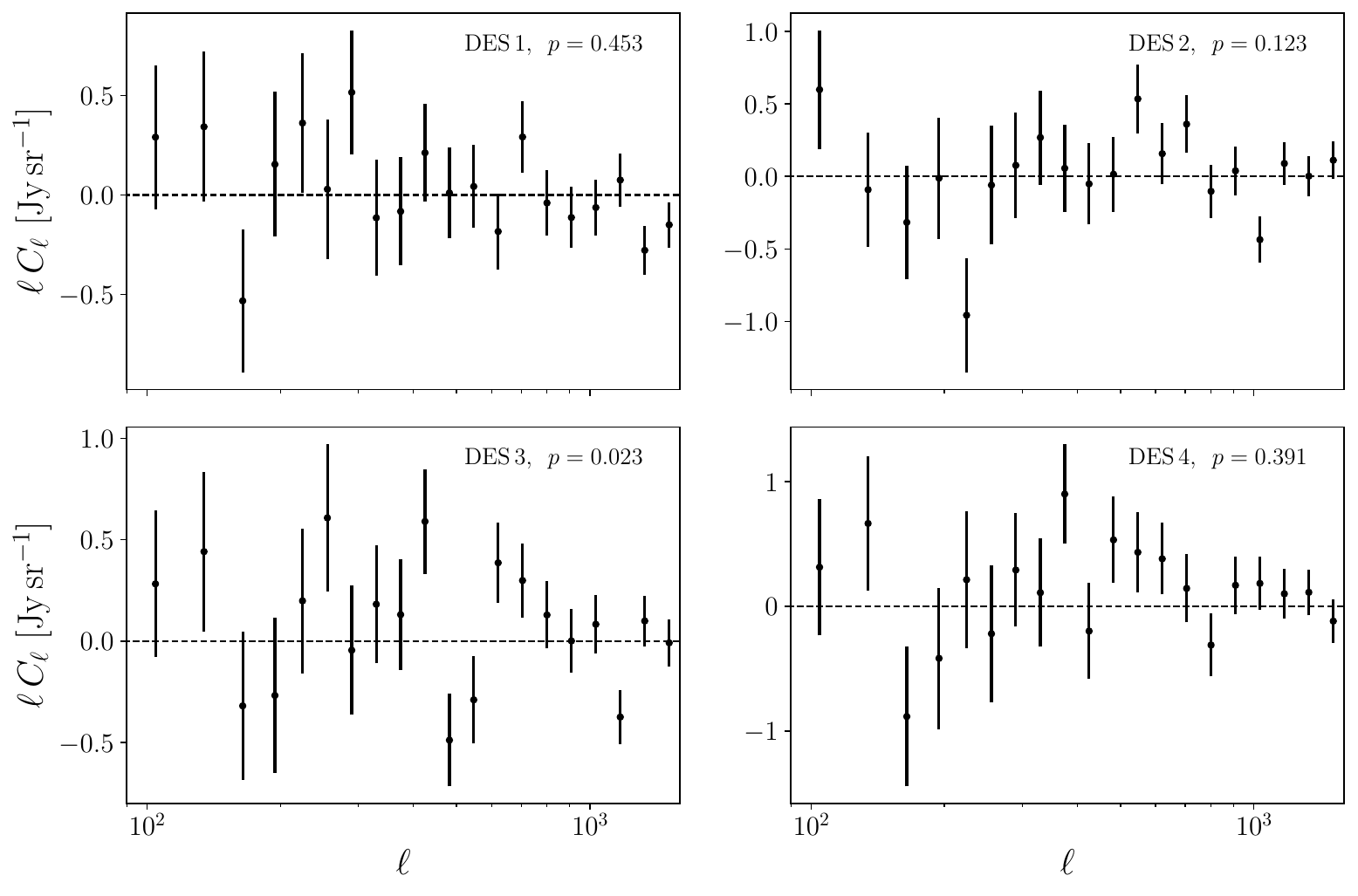}
    \includegraphics[width=0.9\textwidth]{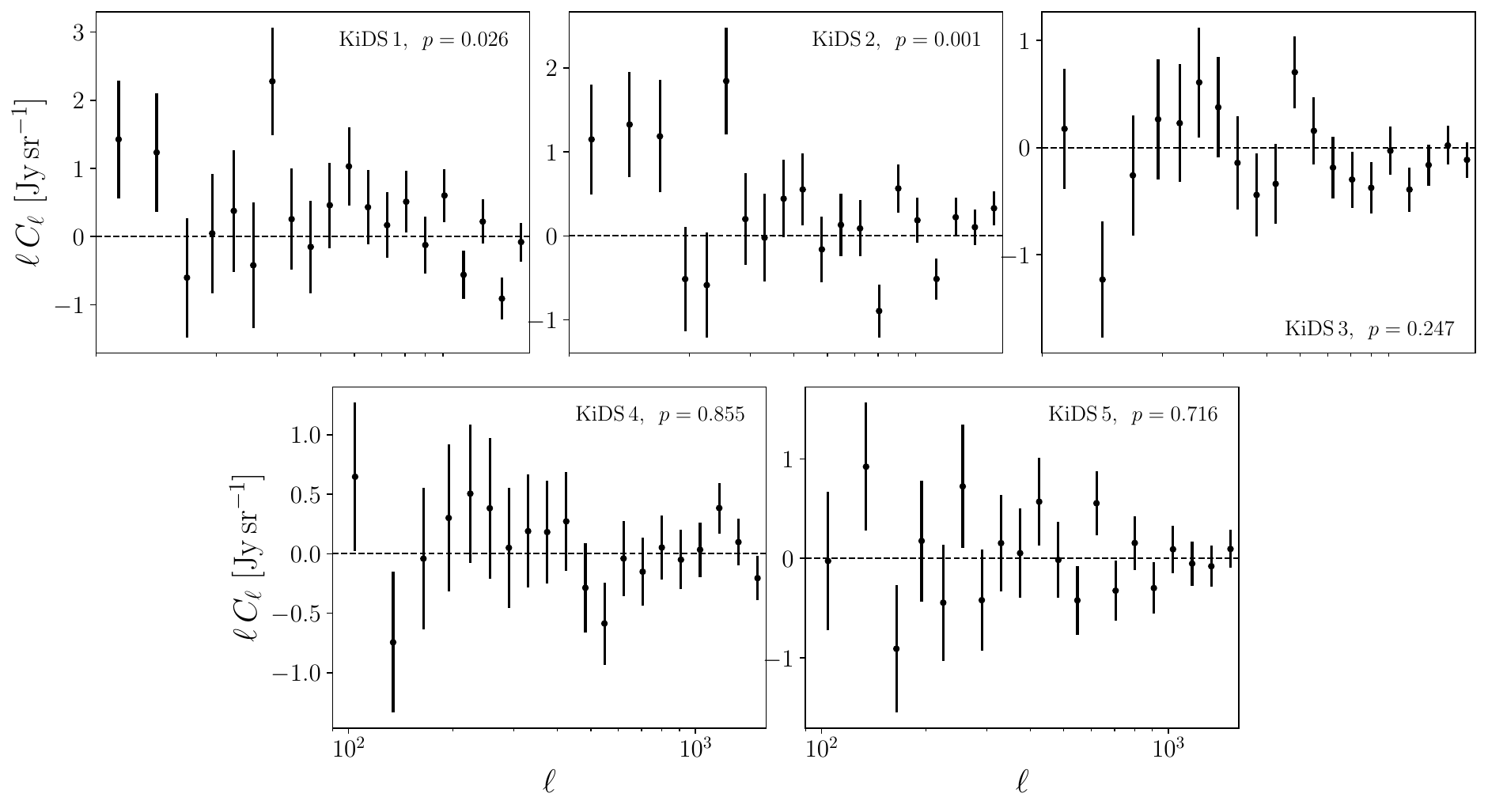}
    \caption{$B$-mode power spectra for the \des and \kids datasets (top and bottom panels). The probability-to-exceed of each power spectrum, corresponding to its $\chi^2$ value with respect to a null signal is included in each panel.}
    \label{fig:b_coadd}
  \end{figure*}
  As an additional test for systematics in the data used here, we have examined the correlations between the shear $B$-modes in each redshift bin of the \des and \kids catalogs, and the coadded CIB map. To quantify the evidence for a potential systematic, we calculate the $\chi^2$ value of each power spectrum with respect to a null signal, and its associated $p$-value. The results are shown in Fig. \ref{fig:b_coadd}. We find overall acceptable $p$-values ($p>2\%$), with the exception of the second \kids bin, for which $p=0.001$. This is in agreement with the findings of the \kids team, which identified evidence of systematics in this bin. As we have shown in the main text, the inclusion of this bin in the analysis has a negligible effect in our final results, since the shear signal at these low redshifts is small.

\bsp	% typesetting comment
\label{lastpage}
\end{document}